\documentclass[12pt,a4paper]{article}
\usepackage{graphicx}
\usepackage{times}
\usepackage{color}

\begin{document}
\sloppy

\title{\bf Superconducting atom chips:\\  advantages and challenges}
\author{V. Dikovsky$^{1}$, V. Sokolovsky$^{2}$,
Bo Zhang$^{3}$, C. Henkel$^{3}$, and R. Folman$^{1,2}$ 
\\[1em]
\centerline{\it $^1$ Ilse Katz Center for Meso- and
Nanoscale Science and Technology,}\\ 
\centerline{\it Ben-Gurion
University of the Negev, Beer Sheva, Israel}\\ 
\centerline{\it $^2$ Department of Physics, Ben-Gurion University of the Negev,
Beer Sheva, Israel}\\ 
\centerline{\it $^3$ Institut f\"{u}r Physik
und Astronomie, Universit\"{a}t Potsdam, Potsdam, Germany}}
\date{21 July 2008}

\maketitle


\begin{abstract}%
    Superconductors are considered in view of applications to atom
    chip devices.  The main features of magnetic traps based on
    superconducting wires in the Meissner and mixed states are
    discussed. The former state may mainly be interesting for
    improved atom optics, while in the latter, cold atoms may provide
    a probe of superconductor phenomena.  The
    properties of a magnetic side guide based on a single
    superconducting strip wire placed in an external magnetic field are
    calculated analytically and numerically.  In the mixed state of
    type II superconductors, inhomogeneous trapped magnetic flux,
    relaxation processes and noise caused by vortex motion are posing
    specific challenges for atom trapping.\\
    PACS: {37.10.Gh} {Atom traps and guides} 
    {85.25.Am} {Superconducting device characterization, 
    design, and modeling}
\end{abstract}


\section{Introduction}

One of the high priority goals in atom chip research is to
increase lifetime and coherence time for ultracold atoms
trapped in magnetic potentials close to the surface. This is
important for both scientific aims and technological applications.
Progress towards this goal demands the control and reduction of
magnetic noise produced by the metallic components of the atom
chip. Randomly fluctuating magnetic fields are generated by
thermal current noise in the conducting chip elements and reduce
the number of trapped atoms (losses), increase their temperature
(heating) and lead to a phase uncertainty in the atom's state
(decoherence) -- see, for example, \cite{Folman,Reichel02,Fortagh07} and references
therein. Theoretical analysis of the magnetic noise generated by a
normal metal \cite{Varpula,Henkel,Sidles,Scheel} predicts a fast
reduction of the lifetime  with the decrease of the distance
$z_{t}$ between the trapped atom and the metal surface (trap
height); this is in excellent agreement with lifetime measurements
\cite{Cornell,Lin,Jones}. At a trap height less than $10\div
20\,\mu{\rm m}$, thermal magnetic noise exceeds all other harmful
influences on the atom cloud (technical noise due to the current
supply instability, residual gas collisions, stray magnetic
fields) and provides the dominant limit for the lifetime
\cite{Folman,Fortagh07}. In the last few years, the application of
superconducting materials to atom chips has been widely discussed
as a perspective to extend the lifetime of cold atoms
\cite{Scheel,Skagerstam,Rekdal1,Skagerstam1,Hohenester}. A recent
theoretical estimate \cite{Skagerstam1} of the magnetic noise
caused by a superconductor in the Meissner state showed that the
lifetime of atoms trapped above a superconducting layer would be,
at least, six orders of magnitude longer than above a normal metal
in the same geometry. The analysis presented in \cite{Hohenester}
predicts an atom lifetime of 5000 s at a trap height of
$1\,\mu{\rm m}$. For comparison, at the same height in a normal
metal trap the lifetime is less than 0.1\,s \cite{Lin}. At larger
heights, the lifetime in a superconducting niobium chip can be
much larger (up to $10^{11}$ s at temperature $T = 4.2\,{\rm K}$).

Two first realizations of atom chips with superconducting elements
have been reported in Refs.\cite{Nirrengarten} and \cite{Casper}.
In both setups, the trapped atoms were $^{87}$Rb. In the
experiment by Nirrengarten et al.\ \cite{Nirrengarten}, the
current-carrying wires (in ``U" and ``Z" shape) were made of
niobium and operated at about $4.2\,{\rm K}$. The obtained atom
spin relaxation time was estimated as 115\,s. This value
is comparable to the best one achieved for atoms trapped near
normal-metal wires \cite{Zhang}. In the second experiment
\cite{Casper}, special efforts have been undertaken to reduce the
influence of technical noise. Using a MgB$_{2}$ film, a
``Z"-shaped wire was fabricated as a part of a closed
superconducting loop and operated in the persistent current
regime. This permits to disconnect the current supply and get rid
of its instability, i.e.\ technical noise. To our knowledge, in both
experiments the trap lifetimes were limited by processes other
than the magnetic noise generated by the superconducting elements
of the atom chip.

Aside from magnetic noise reduction (thermal and technical), 
the application of
superconductors in atom chips may be advantageous due to high
current densities without Joule heating, and practically zero
electric fields across the superconducting elements. 
Atom traps with high currents and strong confinement, as needed for 
some applications, may not be operated in the 
Meissner state because of too low critical parameters. 
One can then use superconducting wires in the so-called mixed state where
magnetic flux partially penetrates the chip structures. This is an issue 
that we address in this paper, complementing previous approaches that focus 
on the Meissner state~\cite{Cano08}. 
It is well known that 
in addition to atom optics, one may use cold atoms as
sensitive probes of current distribution and noise in the nearby
surface of the atom chip~\cite{FolmanScience,Yoni}. As the
mixed state of a superconductor exhibits vortex phenomena and flux
noise much higher than that of the Meissner state, it may be an
interesting object to study in this context, beyond the advantages
of carrying larger super currents. 
The results reported here identify a fairly large parameter
window where atom chips can be designed with superconducting elements,
both in the Meissner and mixed states.

In the following Section, we review typical properties of 
superconducting materials, in particular critical parameters in view
of atom chip applications. Section~\ref{s2:side-guide} analyses side 
guide traps formed by combining a bias field with the supercurrent of
a rectangular wire. We show analytically and numerically how the 
current distribution is significantly modified due to screening and
flux penetration and identify the consequences for trapping and 
transporting atoms on a chip.
We also give an overview on magnetic noise in mixed-state superconductors.
Section~\ref{s4:discussion} discusses 
the confinement of the magnetic trap, compared to normal metal wires.
The analytically solvable case of a cylindrical wire and the method
used for numerical calculations are described in the Appendices.

\section{Typical superconductors}

Let us briefly survey the properties of superconductors
which are important for building a superconducting atom chip.
At low enough fields and temperatures, superconductors exhibit the
Meissner effect: magnetic fields are expelled from
their interior. In this regime, the field
penetrates into a superconductor only over a small depth $\lambda$
from the surface -- the London penetration depth. At $T = 0$,
$\lambda$
is of the order of 50 nm and increases with temperature as
$[ 1-(T/T_{c})^{4} ]^{-1/2}$, where $T_{c}$ is the critical temperature.

Superconducting materials are classified as type I and type II
superconductors that differ in their behavior as the magnetic field
is increased. Type I superconductors (pure metals as Pb,
Al, Hg, In) are in the Meissner state
over the whole sub-critical ranges of temperature $T < T_{c}$,
external magnetic field $B_{0} < B_{c}$, and current $I < I_{c}$.
Their critical parameters are quite low, however, which is the main
problem for applications. The highest values are observed for lead:
$T_{c} = 7.2\,{\rm K}$, critical
magnetic field $B_{c}=0.055\,{\rm T}$ at $T = 0\,{\rm K}$,
and surface (sheet) critical
current density $J_{c} = B_{c} / \mu_{0} = 
4.4\cdot 10^{4}\,{\rm A/m}$
in zero external magnetic field at $T = 0$ \cite{Saint-James}.
($\mu_{0}$ is the free space permeability.)
\begin{table}[tbh]
\caption{\small Critical parameters of selected type-II superconductors.  The
critical current density values are referring to the highest quality
superconducting films and tapes (Nb$_{3}$Sn and
(Pb,Bi)$_{2}$Sr$_{2}$Ca$_{2}$Cu$_{3}$O$_{10}$).
Below the first critical field $B_{c1}$, the material is in the
Meissner state (vortex free), between $B_{c1}$ and $B_{c2}$ in the
mixed state (with vortex penetration).
The temperature is $4.2\,{\rm K}$ unless
otherwise quoted.
Data collected from Refs.\
\cite{Nirrengarten,Finnemore,Takano,Gamaty,%
Yeshurun1,Matsubara,Rudnev,Meng,Larbalestier,Xu}.}
\smallskip\
\begin{small}
\centering
    \begin{tabular}[c]{|l|c|c|c|c|}
        \hline
  Superconductor &{$T_{c}$ (K)}&{$B_{c1}$ (mT)}&{$B_{c2}$ (T)}&{$j_{c}$ ({A/m}$^2$)} \\
        \hline
        \hline
  Nb        & 9.3 & 140  & 0.28  &  $5\cdot10^{10}$\, ({}$B=0$)  \\
        \hline
  Nb$_{3}$Sn & 18 & 40  & 27 & $6\cdot10^{10}$\,
               ({}$B=1\,{\rm T}$) \\
        \hline
  MgB$_{2}$ & 39 & 30   & 15  & $3.5\cdot10^{11}$ ({}$B=0$)  \\
   \hline
  YBa$_{2}$Cu$_{3}$O$_{7-\delta}$& 92 & 25 (${B}\Vert ab$,
  $T \rightarrow 0$)
    & $>$100 (77\,K)& $7.2\cdot10^{11}\,$ ({}$B=0$) \\
    (YBCO)& & 90 (${B}\Vert c$, $T \rightarrow 0$) & & \\
     \hline
    (Pb-Bi)$_{2}$Sr$_{2}$Ca$_{2}$Cu$_{3}$O$_{10}$ & 108 & 13
                 (${B}\Vert ab$, $T \rightarrow 0$)
    & $>$100 (77\,K)& $\approx10^{10}$ ({}$B=0$) \\
     & & & & \\
    \hline
    \end{tabular}
\end{small}
\label{t:sc-table}
\end{table}

Much higher critical parameters are observed in type II
superconductors. We give typical examples of this type in
Table~\ref{t:sc-table}, including niobium, its compound
Nb$_{3}$Sn, as well as the high-temperature superconductors (HTSC)
YBa$_{2}$Cu$_{3}$O$_{7-\delta}$ and
(Pb-Bi)$_{2}$Ca$_{2}$Sr$_{2}$Cu$_{3}$O$_{10}$. These materials
exhibit the Meissner effect below the lower critical field, $B <
B_{c1}$, which is relatively small ($50 \div 500$\,G at 0\,K). The
largest value at $T = 0$ is found for niobium ($B_{c1}(0) \approx
170\,{\rm mT}$) \cite{Finnemore}. With increasing temperature, the
lower critical field falls down approximately as $B_{c1}(0) [1 -
(T/T_{c})^2]$, and other parameters ($B_{c2}$, $j_{c}$) also
decrease towards zero as $T \to T_{c}$. The details of these laws
depend on the type of superconductor and its
fabrication~\cite{Saint-James}. In Table~\ref{t:sc-table}, we
present main critical parameters at $T = 4.2\,{\rm K}$. For highly
anisotropic HTSC with layered crystal structures, the lower
critical field (as well as the critical current density) depends
on the direction of the field relative to the
$ab$-planes~\cite{Ginsberg}. In magnetic fields (external or
caused by a transport current) higher than $B_{c1}$, the magnetic
flux penetrates into a type II superconductor in the form of
vortices that arrange into a more or less regular flux-line
lattice (Abrikosov lattice) which is pinned by inhomogeneities of
the material \cite{Saint-James}. Each vortex carries one quantum
of magnetic flux $\Phi_{0} = \pi \hbar / e = 2.07\cdot 10^{-15}$ T
m$^{2}$ ($e$ is the electron charge and $\hbar$ the Planck
constant). In the simplest (isotropic) case, it looks like a cylindrical
tube of radius $\sim\lambda$ in which superconducting screening
currents circulate around a normally conducting core of radius
$\sim \xi$ (superconductor coherence length).  Both $\lambda$ and
$\xi$ depend on temperature, with $\lambda \gg \xi$ for most type
II superconductors. The mixed state is observed up to the upper
critical field $B_{c2}$ where the vortex cores merge and the
superconducting state is destroyed.

If a superconducting material is used in atom chips, it is
particularly important that the lower critical field $B_{c1}$ and
the critical current density $j_{c}$ be large. The second critical
field $B_{c2}$ for most type II superconductors is typically too
large to be a relevant limit in the magnetic fields applied in
usual atom chip setups. Both lower critical field and critical
current density are ``technology dependent'' because they are very
sensitive to crystal defects (for details see
Refs.\cite{Larbalestier,Blatter,Kossowsky}). The $j_{c}$ values
collected in Table~\ref{t:sc-table} refer to films and tapes of
the best quality.  For example, in the atom chip experiments of
Refs.\cite{Nirrengarten,Casper}, the critical current density was
$5\cdot 10^{10}$ A/m$^{2}$ (niobium film) and $10^{11}$ A/m$^{2}$
(MgB$_{2}$ film) respectively.

One specific property, which distinguishes type II superconductors
from both normal metals and superconductors in the Meissner state,
is their capability to freeze a magnetic flux \cite{Saint-James}.
This effect is due to pinning forces. The vortices can move under
the action of the transport current, when the Lorentz force $j
\Phi_{0}$ is stronger than the pinning force, which can be
estimated as $j_{c}\Phi_{0}$ where $j_{c}$ is the critical current
density. This flux motion results in energy dissipation and
induces a voltage drop along the superconducting element. A
voltage drop of $1\,\mu$V/cm is usually taken as a criterion to
define the critical current. Another mechanism of energy
dissipation predicted by P. Anderson operates at sub-critical
currents and is connected with thermally activated jumps of
vortices out of pinning centres, which also generate electric
fields \cite{Anderson}. The motion of vortices under various
conditions has been investigated by many authors, see for example
Ref.\cite{Blatter,Kossowsky}.

\section{Side guide traps with superconducting wires}
\label{s2:side-guide}

\subsection{Rectangular wire in the Meissner state}
\label{s2:Meissner-state}

We consider magnetic traps in the ``side guide'' configuration to
illustrate the differences between normal metal and
superconducting chips. In this section we analyse a trap based on
wires having the form of a thin strip which is the usual shape in
present day atom chips.  The case of cylindrical conductors is
considered in Appendix~\ref{a:cylinder}.

\subsubsection{Horizontal bias field}

\begin{figure}[tbp]
   \centering
   \hspace*{-10mm}
   \parbox[c]{0.7\textwidth}{
   \includegraphics[width=0.8\textwidth]{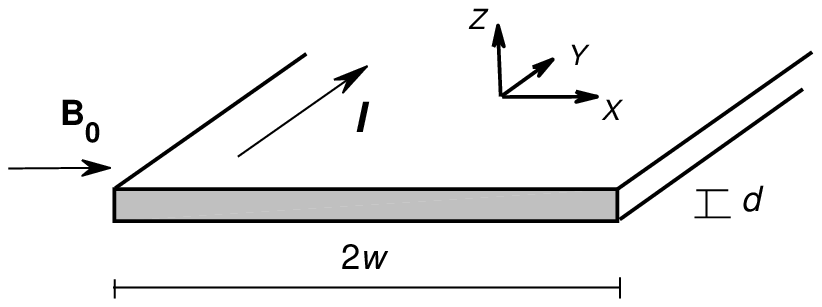}
   }   \hspace*{-05mm}
   \parbox[c]{0.3\textwidth}{
   \includegraphics[width=0.3\textwidth]{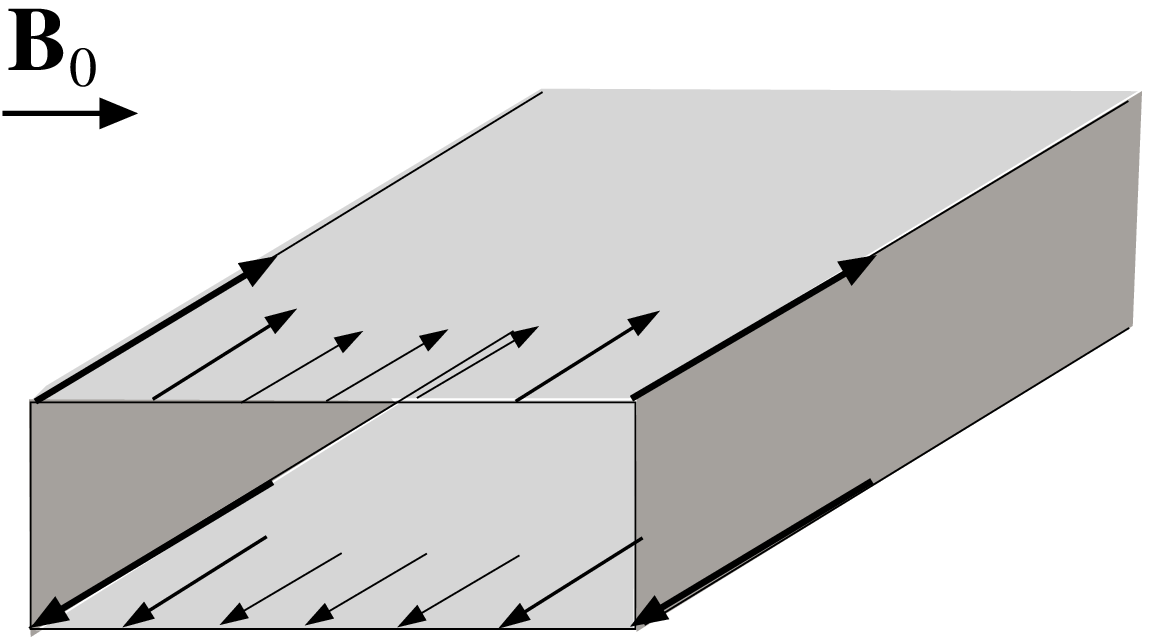}
   \includegraphics[width=0.3\textwidth]{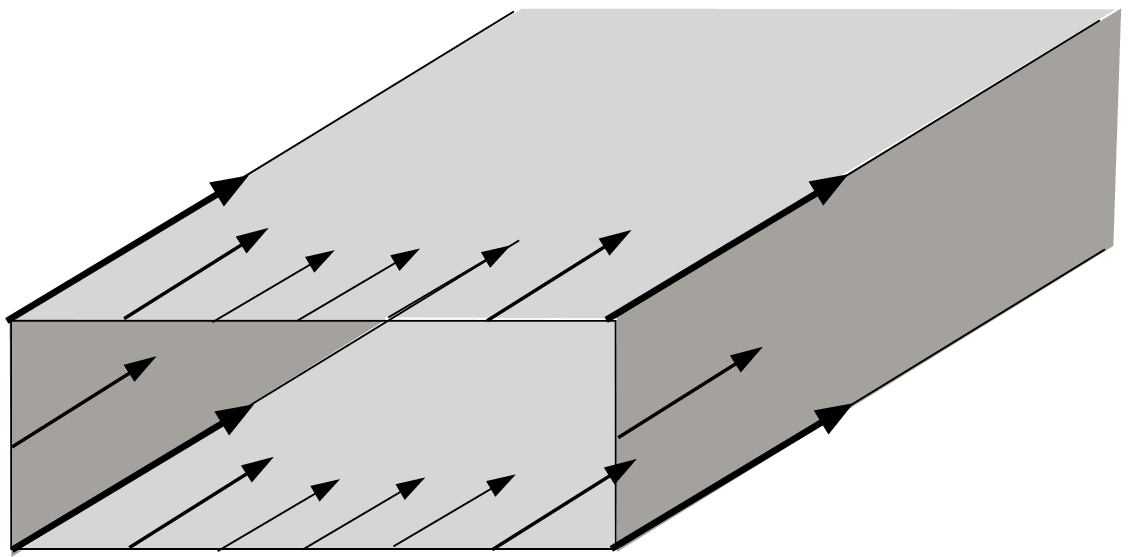}
   }
   \caption{\small{}(left) Superconducting strip carrying a current and placed
   in an external magnetic field. $z=0$ is at the top surface of the strip.
   (right) Illustration of the surface
   current distribution (arrows) for the strip wire.
   Top: screening supercurrents in an external field ${\bf B}_{0}$,
   parallel to the top wire surface (this direction is called ``horizontal").
   The screening currents are particularly large at the corners, but have opposite
   directions on the top and bottom faces; the total current is zero.
   Bottom: transport current without an external field. The current
   density is maximal near the corners [see also
   Eq.(\ref{eq:sheet-current})].
   }
   \label{fig1:sketch-strip}
\end{figure}

Let us start with a superconducting strip wire in the Meissner state.  
The strip is infinitely long along the $y$-axis, with a width $2w$ along
$x$ and a thickness $d$ along $z$, perpendicular to the chip surface
(Fig.\ref{fig1:sketch-strip}).
The strip is placed in an external bias field ${\bf B}_{0}$ parallel
to the $x$-axis.
If we assume that the strip is thin, $d \ll w$, the real current
density distribution can be replaced by a sheet current
$J( x )$ (in ${\rm A/m}$) determined by
integrating the current density over the strip thickness.  The
conditions under which these assumptions are fulfilled are discussed in
\cite{Brandt,Zeldov}.
Regarding the
magnetic field around a thin superconducting strip, one expects that
the $z$-dependence of the current density can be neglected if the
thickness is much less than the observation distance, $z \gg d$.  The
thin strip approximation is therefore expected to be valid
for superconducting films
with typical thicknesses of a few hundred nanometers and
trap heights above several microns.
We confirm this expectation by comparing to numerical
calculations that take into account a finite
thickness.\footnote{%
These calculations are outlined in Appendix B.  For the
sake of faster convergence and to avoid singular fields,
we have `rounded' the edges of the wires
with a radius of curvature $r \approx w / 32$.}

We calculate first the current distribution in the strip
and then the outer magnetic field using the Bio-Savart law.
In this section, we focus on strips in the Meissner state where
magnetic flux does not penetrate.
Any magnetic field perpendicular to the strip surface is completely
shielded by appropriate screening currents. 
If a bias field ${\bf B}_{0}$ is
applied parallel to the wide strip surface, there are no screening
currents in an infinitely thin strip, and the sheet current is given
by the transport current alone.
This situation is changed when the finite thickness is taken into
account.

\begin{figure}[tbp]
    \centering
   \includegraphics[width=67mm]{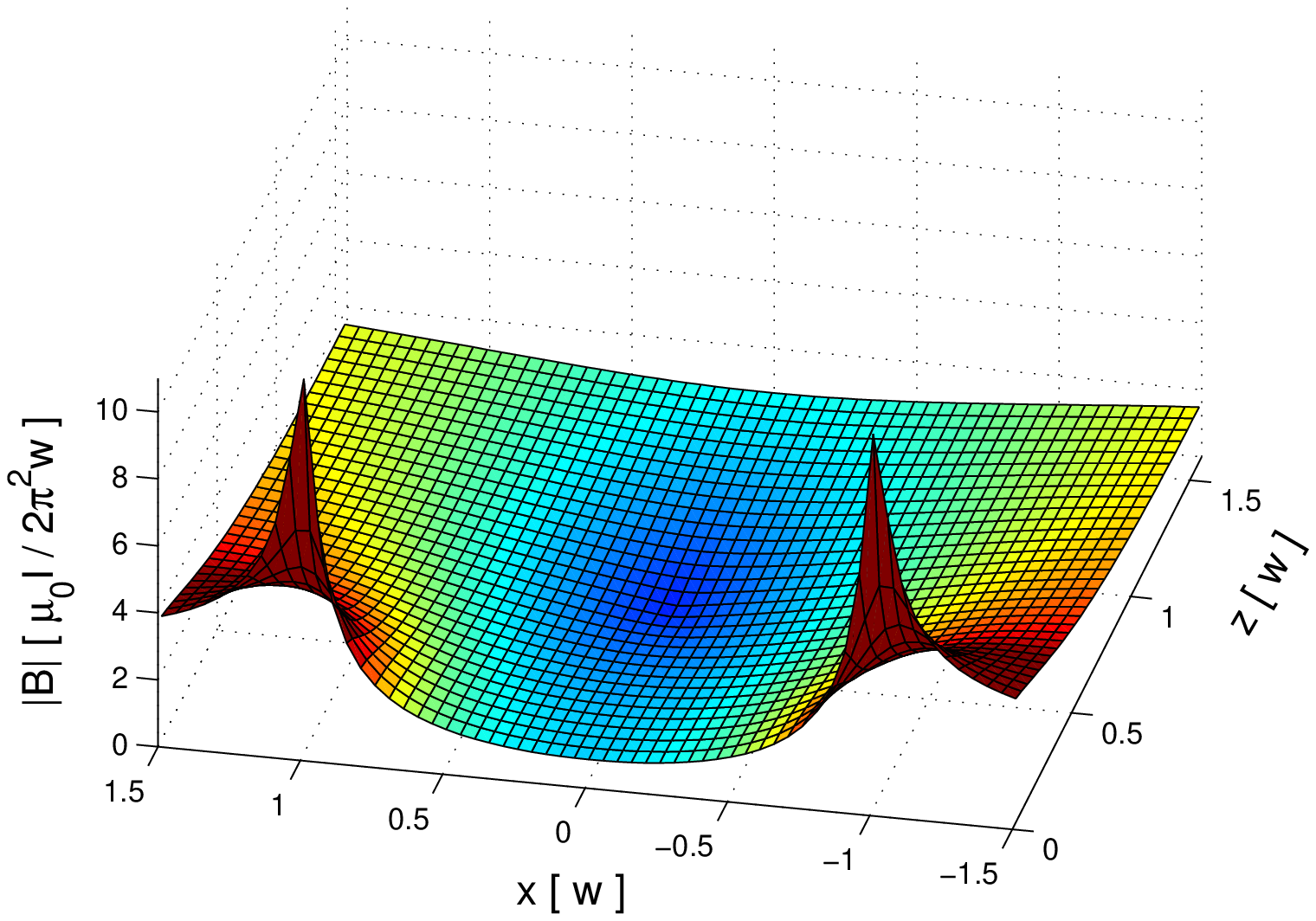}
   \includegraphics[width=67mm]{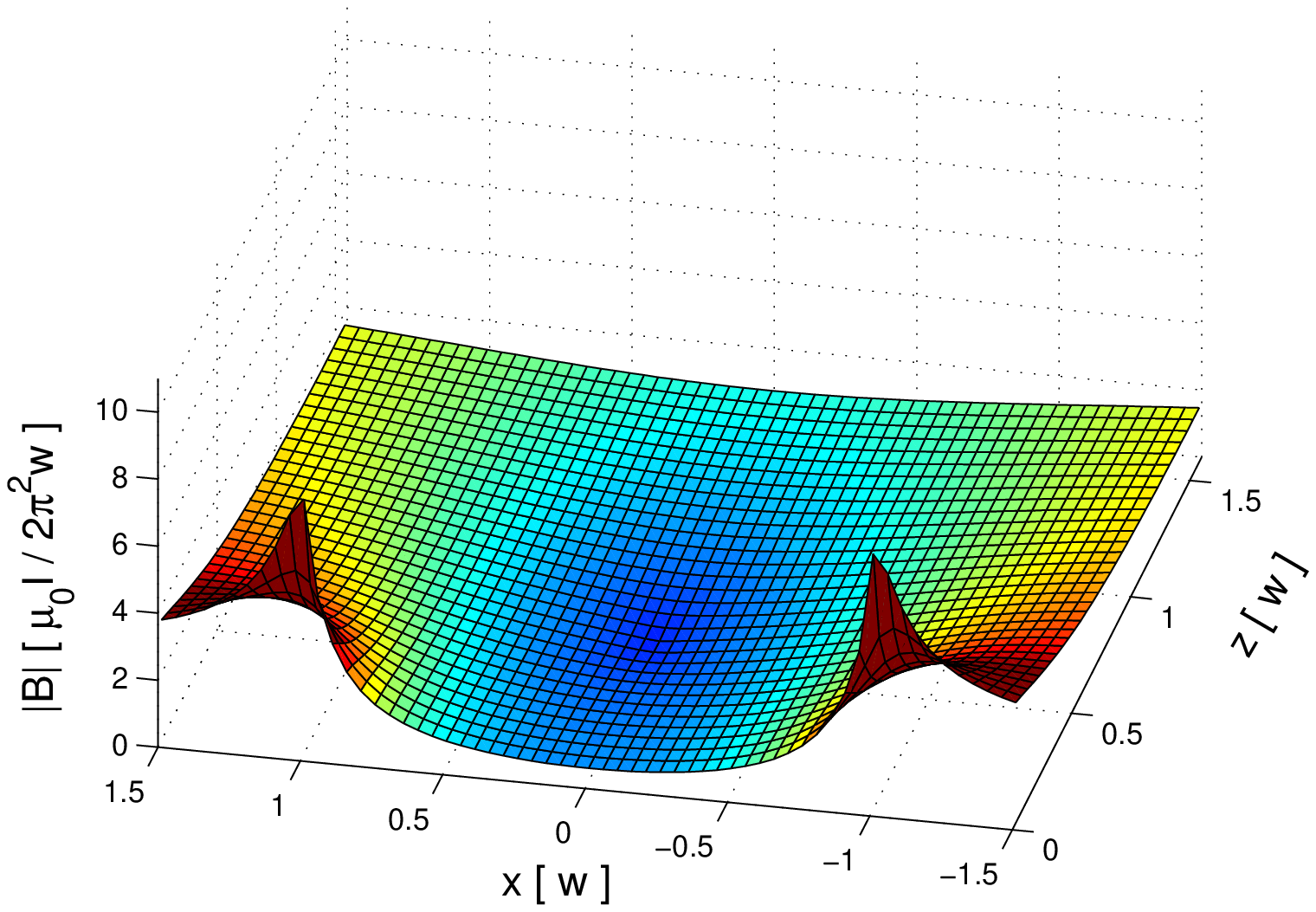}
    \includegraphics[width=65mm]{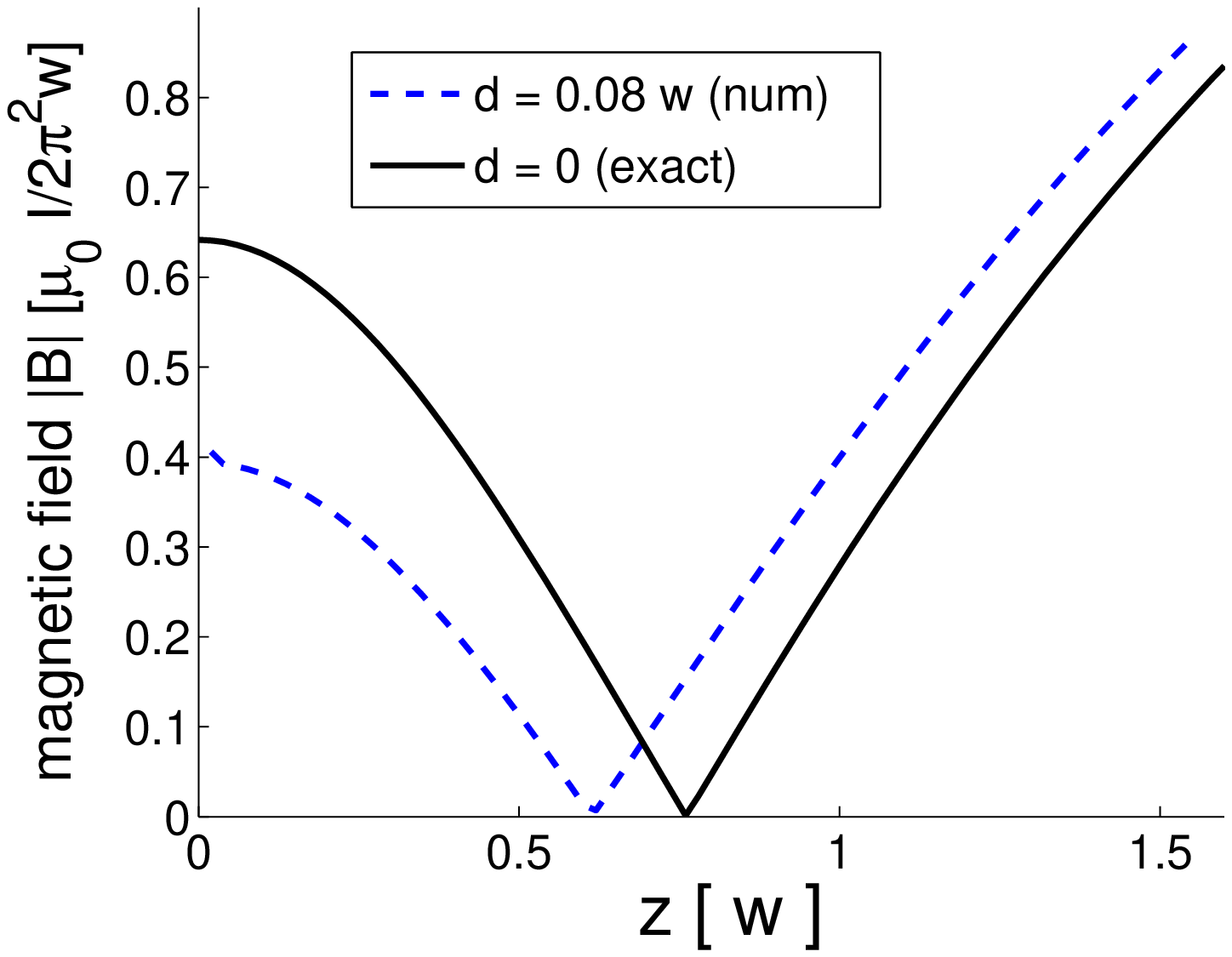}
    \includegraphics[width=65mm]{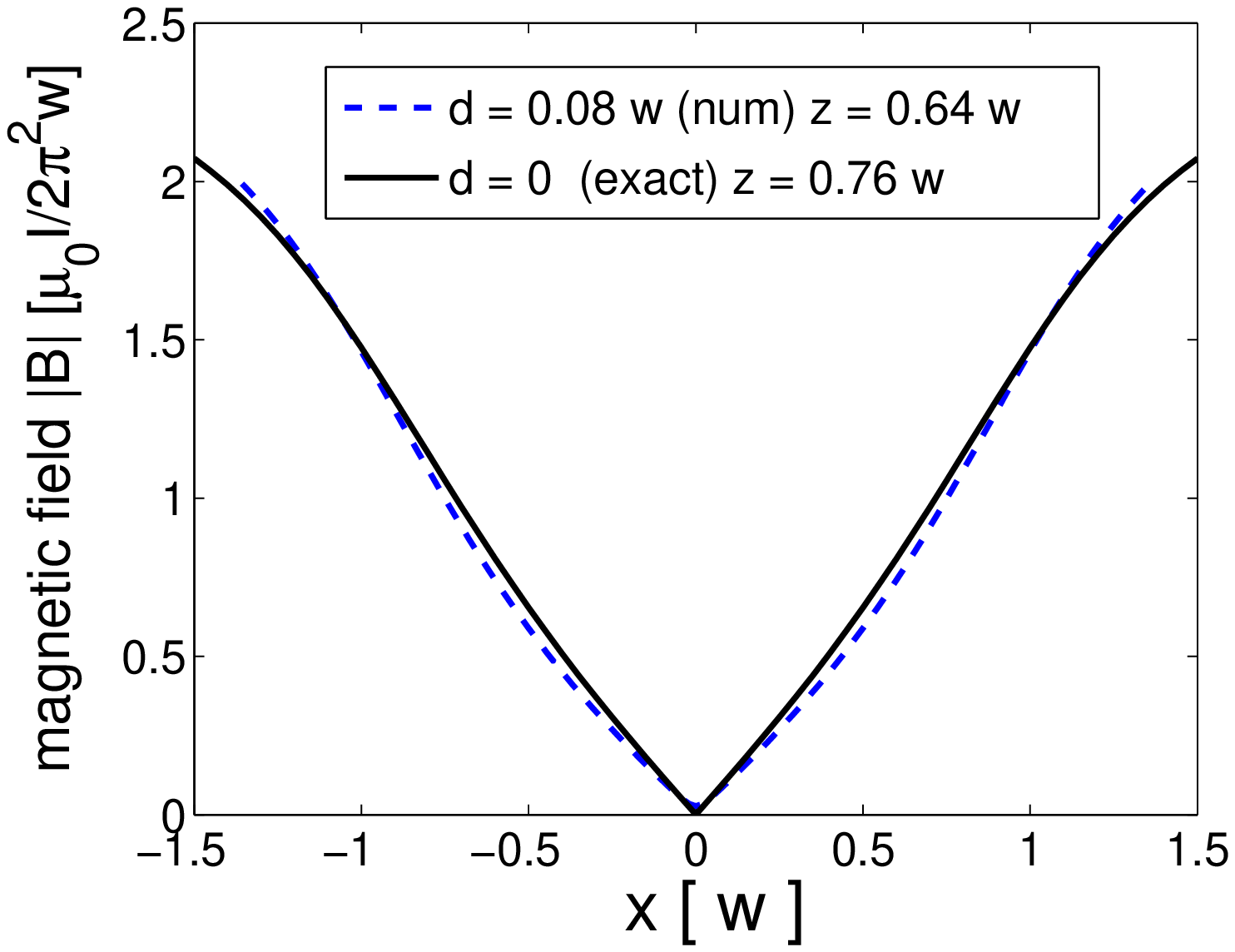}
    \caption[]{\small{}(top) Magnetic trap (``side guide'') created by a
    superconducting rectangular wire with a transport current, placed
    in a horizontal bias field (as shown in Fig.\ref{fig1:sketch-strip}).
    We plot the modulus of the magnetic field in units of
    $\mu_{0} I/(2\pi^2 w)$.  For $I = 1\,{\rm A}$ and $w = 1\,{\rm
    mm}$, this unit corresponds to $(2/\pi)\,{\rm G}$.
    (top left) infinitely
    thin wire (analytical calculation); (top right)
    finite thickness $d \approx 0.04\times2w$ (numerical calculation).
    (bottom)
    Cross-sections along the $z$-axis and $x$-axis through the trap 
    centre ($x=0$ is chosen in the middle of the top wire face.)
    Solid lines: analytical results for an infinitely thin
    strip in the sheet current approximation; dashed lines: numerical
    calculation for a finite thickness.
    Due to the finite thickness, the trap is shifted closer
    to the wire surface (compared to the case $d=0$ the trap height
    is reduced from $z_t = 0.76\,w$ to $0.64\,w$). Half of this 
    shift can be explained by measuring the distance from the center 
    of the wire, and shifting the dashed curve by $d/2$ would make it
    coincide with the solid curve at large distances.
    The bias field is $B_{0} = -2.5 \,\mu_{0} I / (2\pi^2 w)$. The
    numerical calculations use a thickness $d \approx 0.04\times 2w$,
    with rounded corners (radius $0.031\,w$).
    }
\label{fig:b-strip}
\end{figure}

\begin{figure}[tbh]
    \centering
    \includegraphics[width=65mm]{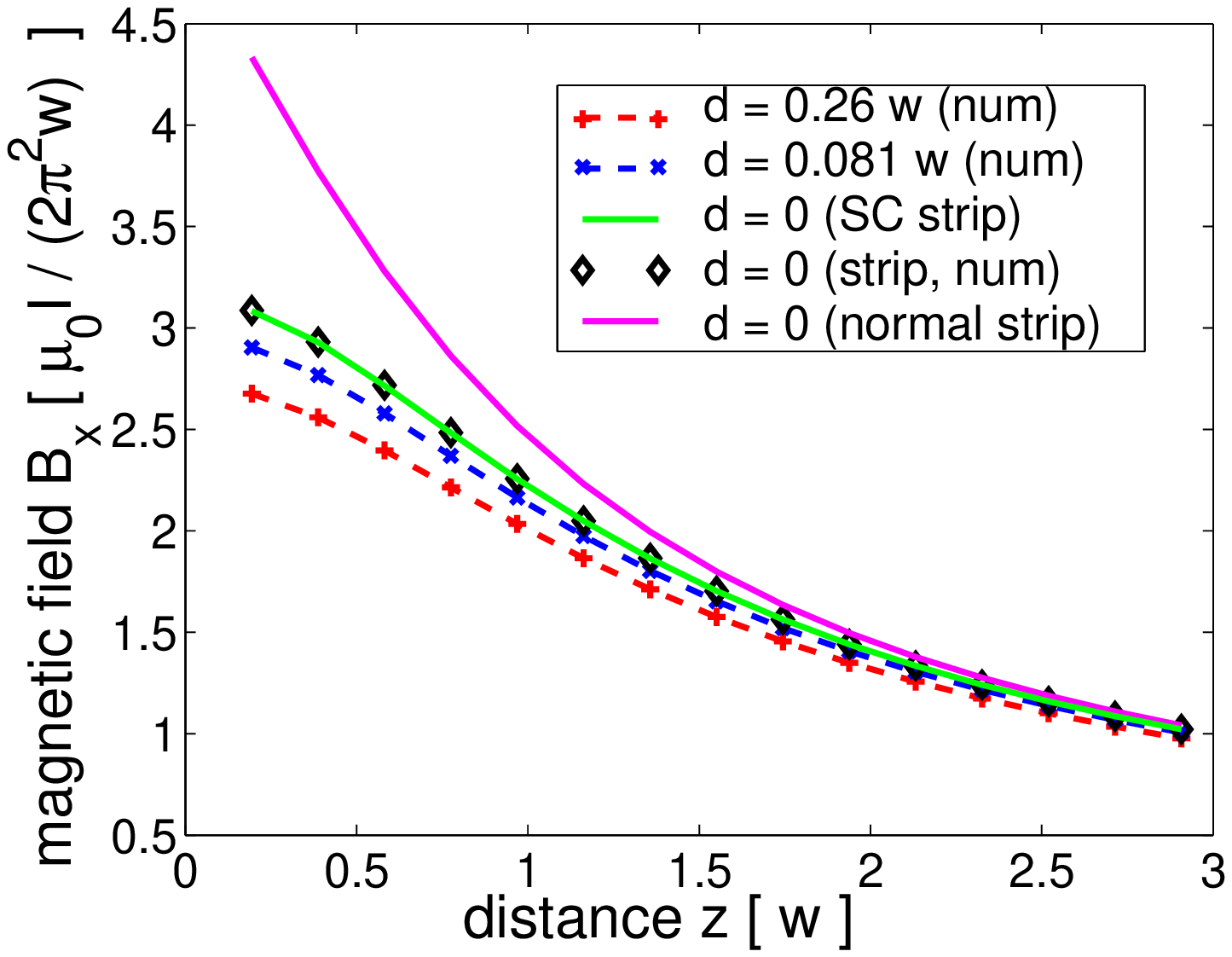}
    \includegraphics[width=65mm]{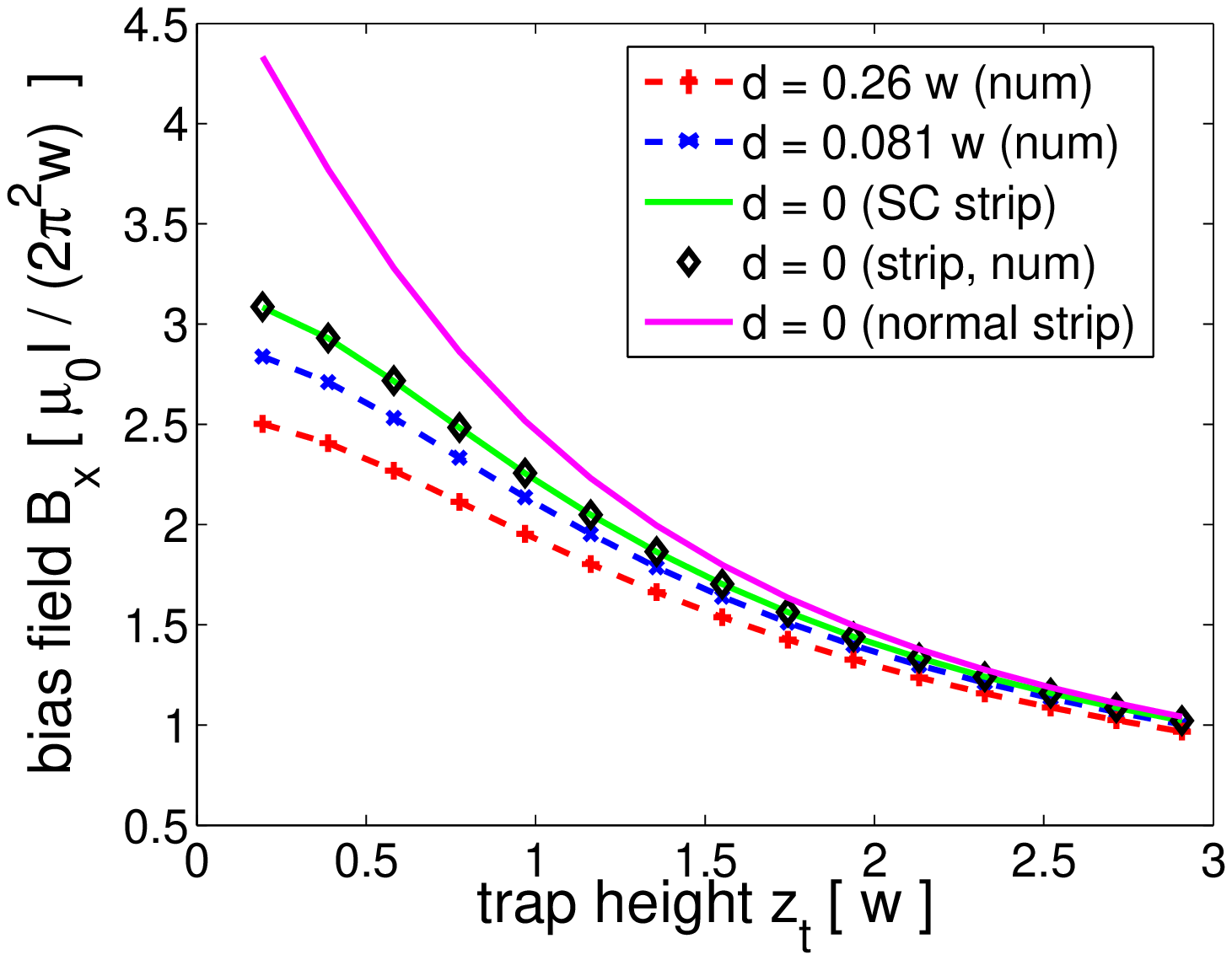}
\caption{\small{}(left) Magnetic field created by a supercurrent $I$
in a rectangular wire as a function of the observation distance $z$
(in units of the strip half-width $w$, measured from the wire top 
surface).  Solid lines
give analytical results for $d=0$: magenta -- normal metal strip, green --
superconducting strip (superimposed diamonds: numerical calculation).
Dashed lines (with symbols) give numerical results for
superconducting strip of different thickness: blue --
$d=0.04\times 2 w$, and red -- $d=0.13\times 2 w$.
(right) Bias field required to form a trap at distance $z_{t}$ above
a strip wire.  This plot differs from the left one only for wires of
finite thickness where screening currents
caused by the bias field are different on the top and bottom
surfaces. For this case, the trap height is slightly reduced.
}
    \label{fig:trap-height}
\end{figure}

For an infinitely thin strip, the sheet current profile along the
$x$-axis as well as the field distribution around the strip can be
calculated
analytically. The sheet current density is
given by \cite{Brandt,Zeldov}
\begin{equation}
    J( x ) = \frac{ I / \pi }{
    \sqrt{ w^2 - x^2 } }
    \label{eq:sheet-current}
\end{equation}
The magnetic field above the strip is presented in
Fig.\ref{fig:b-strip}(top).
We actually plot its modulus $|{\bf B}( {\bf r} )|$ since this
provides the magnetic trapping potential in the adiabatic
approximation.
Due to symmetry,
the minimum modulus of the total field occurs on the $z$-axis (above
the centre of the strip) where the field caused by the transport
current $I$ is parallel to the $x$-axis:
\begin{equation}
    B_{x}( x=0, z ) =
    \frac{ \mu_{0} I z } { 2 \pi^2 }
    \int\limits_{-w}^{w}\!\frac{ {\rm d}x' }{
    \sqrt{ w^2 - {x'}^2 } ( {x'}^2 + z^2 ) }
    =
    \frac{ \mu_{0} I } { 2 \pi \sqrt{w^2 + z^2} }
    \label{eq:Hx-strip-above-center}
\end{equation}
This can be compared to the field above a normally conducting strip.
Here, the current distribution can be taken as uniform, and the
Biot-Savart law gives \cite{Reichel02}
\begin{equation}
    \mbox{normal wire: }
    B_{x}( x=0, z ) =
    \frac{ \mu_{0} I } { 2 \pi w }
    \arctan( w / z )
    \label{eq:Hx-strip-above-center-normal}
\end{equation}
The field profiles, described by (\ref{eq:Hx-strip-above-center})
and (\ref{eq:Hx-strip-above-center-normal}) as well as the
numerical results for two
thickness/width ratios are plotted in Fig.\ref{fig:trap-height}. By
adding
a homogeneous bias field along the $x$-axis with value $B_{0} =
- B_{x}( x=0, z_{t} )$, a magnetic quadrupole trap is formed at
height
$z_{t}$.
We see that for a superconducting strip, the required bias field is
smaller than
for a normal conductor, by a factor $2/\pi$ at small trap height.
This is a result of the different
distribution of current density. Both wires behave practically the same above
heights $z \ge 2.5\,w$. Similar results are obtained for a cylindrical
wire (see Appendix~\ref{a:cylinder}).

Figs.\ref{fig:b-strip}, \ref{fig:trap-height} also illustrate the
impact of the finite thickness of the superconducting wire.  The bias
field induces screening currents on the top and bottom
faces of the wire
and makes the currents on the top and bottom sides differ
(Fig.\ref{fig1:sketch-strip}, top right).
The effect is, however,
quite small for the parameters that we explored (up to thickness $d
\approx w / 4$).  For example, the magnetic field gradient in the trap
centre is practically unchanged (Fig.\ref{fig:b-strip}).  The main
tendency is to bring the trap position closer to the wire surface (at
fixed bias field), see Fig.\ref{fig:b-strip}.  Conversely, the minimum
bias field required to create a trap at nonzero height is lowered
(Fig.\ref{fig:trap-height}, right).

\subsubsection{Vertical bias field}

The influence of a magnetic field on the current
distribution in a flat superconducting wire is most pronounced
when the field is perpendicular to the wide surface of the wire.
For this reason, different loading procedures have to be designed
for superconducting atom chips compared to normal ones.
Typically, a chip trap is loaded by transporting
(positioning) an atom cloud along the $x$-axis adjusting bias
fields and currents.
In a normal-metal atom chip this is performed using a vertical bias
$B_{z}$. Above a superconducting wire, the procedure must be altered
as the wire builds up significant
screening currents to shield its interior from a vertical field.
The field profile above the chip becomes significantly non-uniform,
leading to potential barriers that have to be taken into account for
the cloud transport.
This effect is maximal for superconductors in the Meissner state and
provides the most striking difference to a normal conductor.

\begin{figure}[htbp]
    \includegraphics [width=0.47\textwidth]{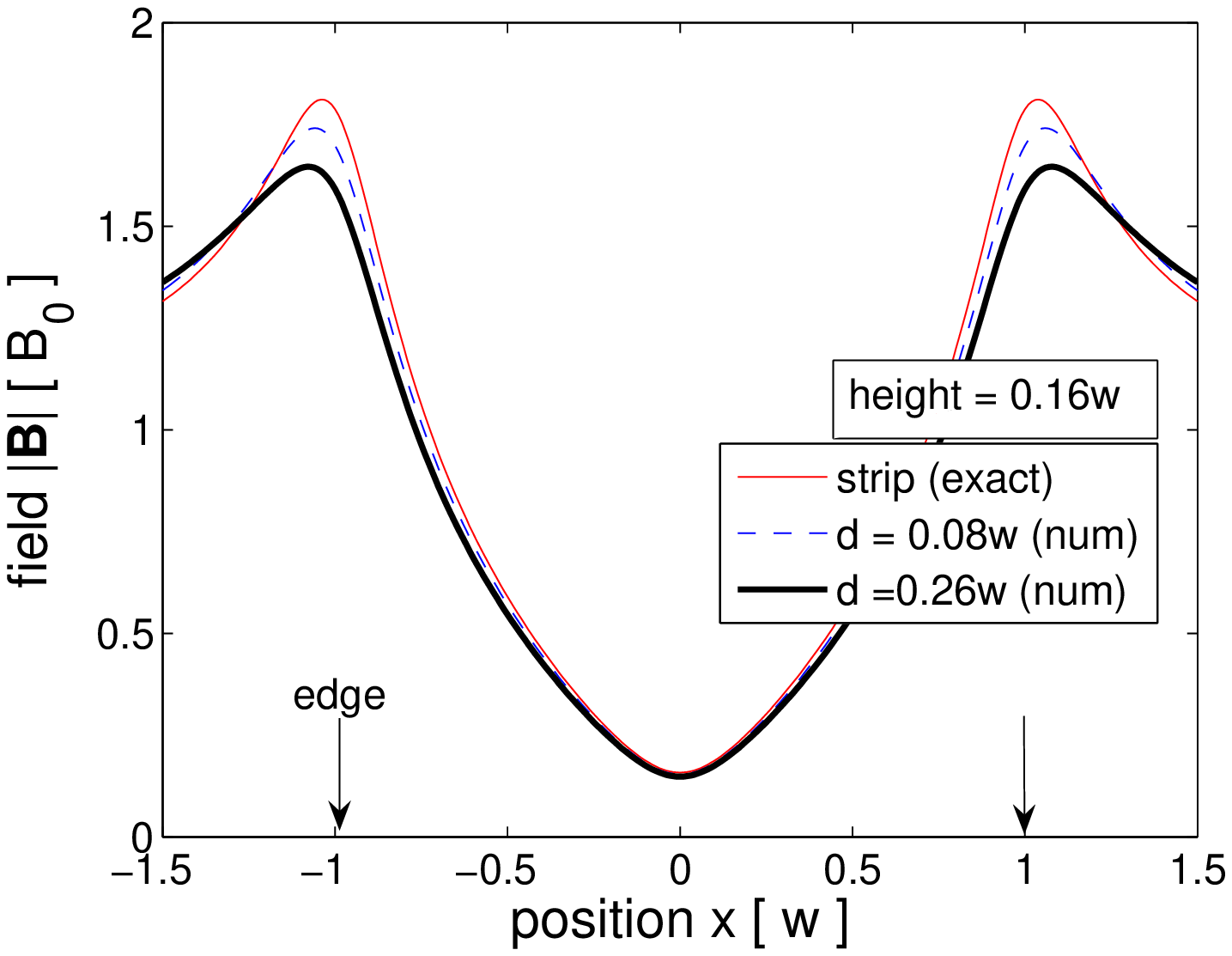}
\hfill
    \includegraphics [width=0.47\textwidth]{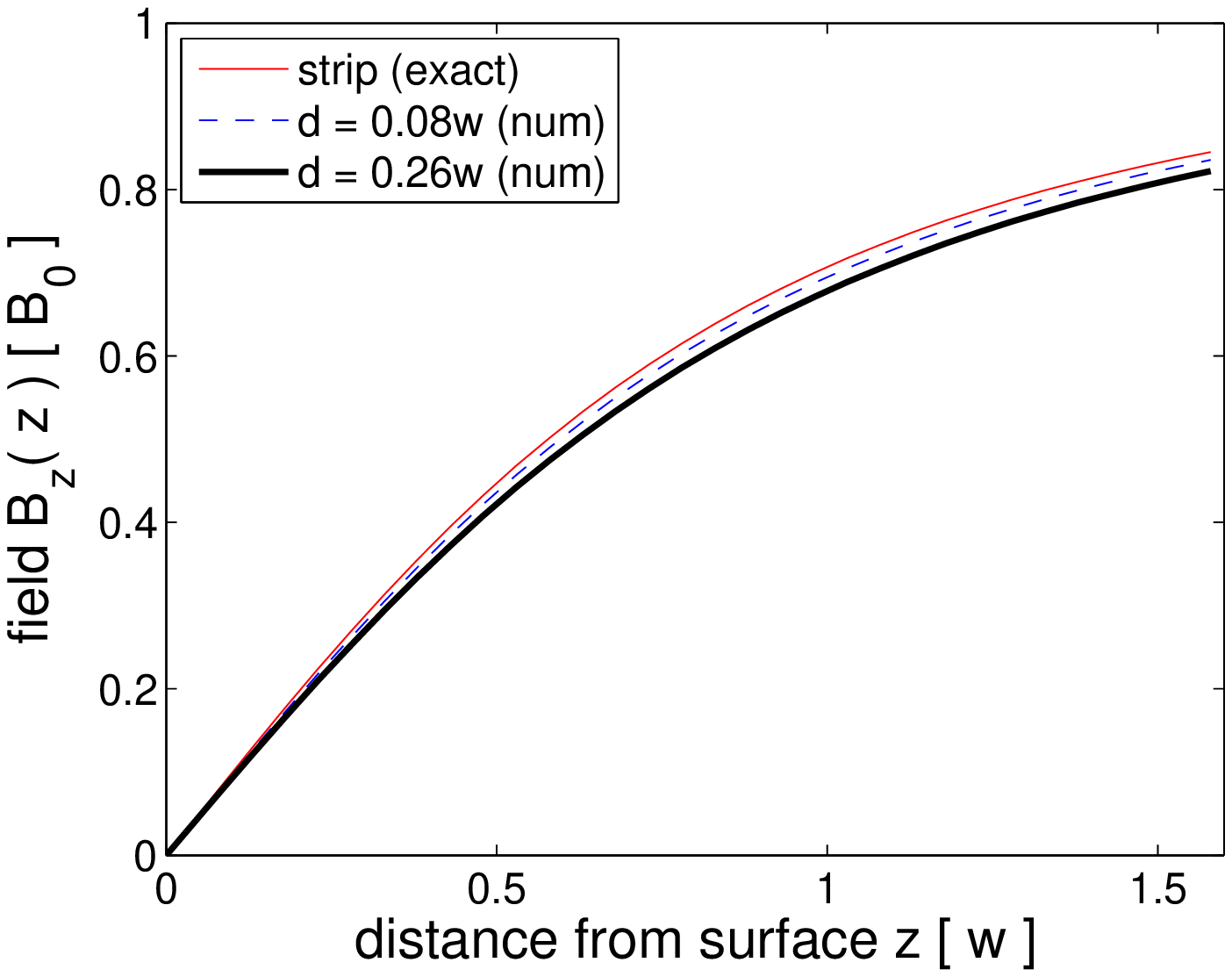}
\caption{\small{}Modulus of magnetic field above a thin
superconducting strip placed in a vertical magnetic field of
magnitude $B_{0}$.
(left)
Field at fixed height, $|{\bf B}( x, z = 0.16\,w)|$,
given in the units of the applied field $B_{0}$. (right) Field
$|{\bf B}( x = 0, z )|$. The position $x = 0$ corresponds to the
strip centre above which the field is oriented perpendicular to
the strip; due to symmetry and continuity, it must go to zero at
the wire surface. The lines compare calculations for strips of
different thickness $d$. Infinitely thin strip: analytical
calculation based on Ref.\cite{Brandt}; $d > 0$: numerical
calculations. } \label{Fig4}
\end{figure}

As an example, we illustrate in Fig.\ref{Fig4} the magnetic field
above a superconducting strip placed in a vertical bias. Magnetic
field maxima occur at the edges of the strip
(Fig.\ref{Fig4}(left)) that exceed significantly the applied bias
field near the wire surface. This field increase is reduced to
about 15\% at a height of $1.5\,w$, but should be taken into
account for loading the trap. Note that in order to maintain the
Meissner state, the total magnetic field should be less than the
lower critical field $B_{c1}$ in any point of the superconductor
surface. The vertical magnetic field near the edges of the
superconducting strip increases proportionally to the ratio
width/thickness ($2w/d$) and may become significant
for a wide strip~\cite{Landau}. The same magnetic field increase near the strip
edges is induced by the transport current.\footnote{This is a
rough approximation which takes into account only the
demagnetisation factor. A more accurate analysis is given, for
example, in \cite{Kuznetsov}.} This field concentration can result
in the partial transition of the sample into the mixed state even
if the current and magnetic fields are far from the critical
values. Magnetic traps based on superconducting wires in the mixed
states are discussed in the next section.

\subsection{Superconducting guiding wire in the mixed state}
\label{s3:mixed-state}

\subsubsection{Side guide}

\begin{figure}[tbp]
\includegraphics*[height=50mm]{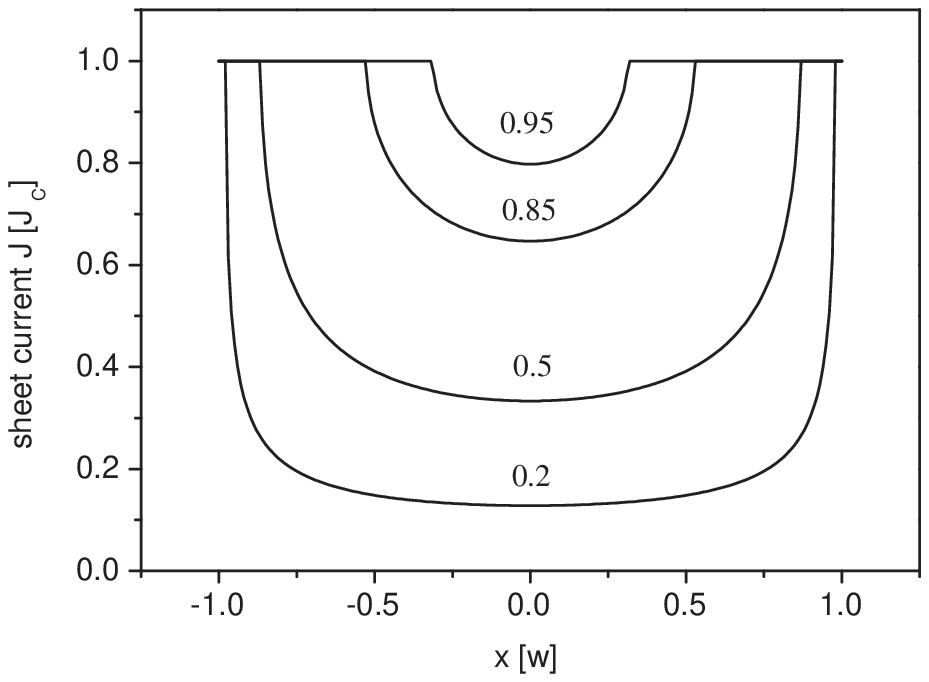}
\includegraphics*[height=58mm]{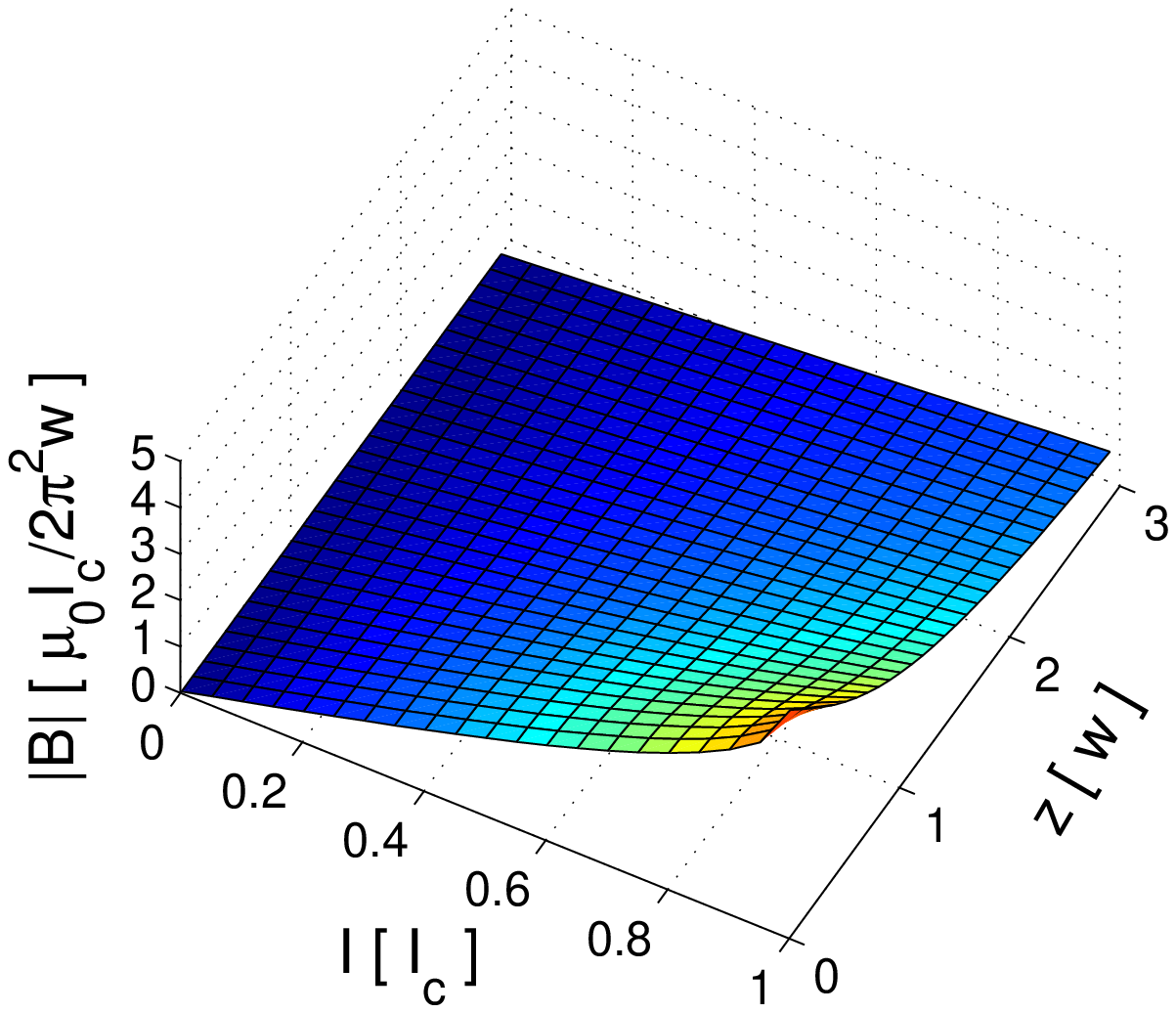}
\includegraphics*[height=49mm]{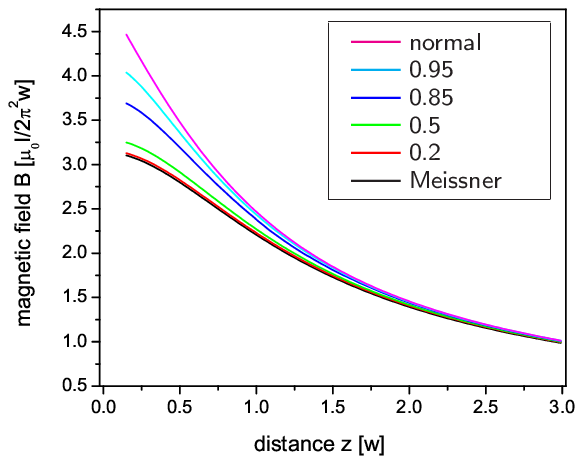}
\includegraphics*[height=50mm]{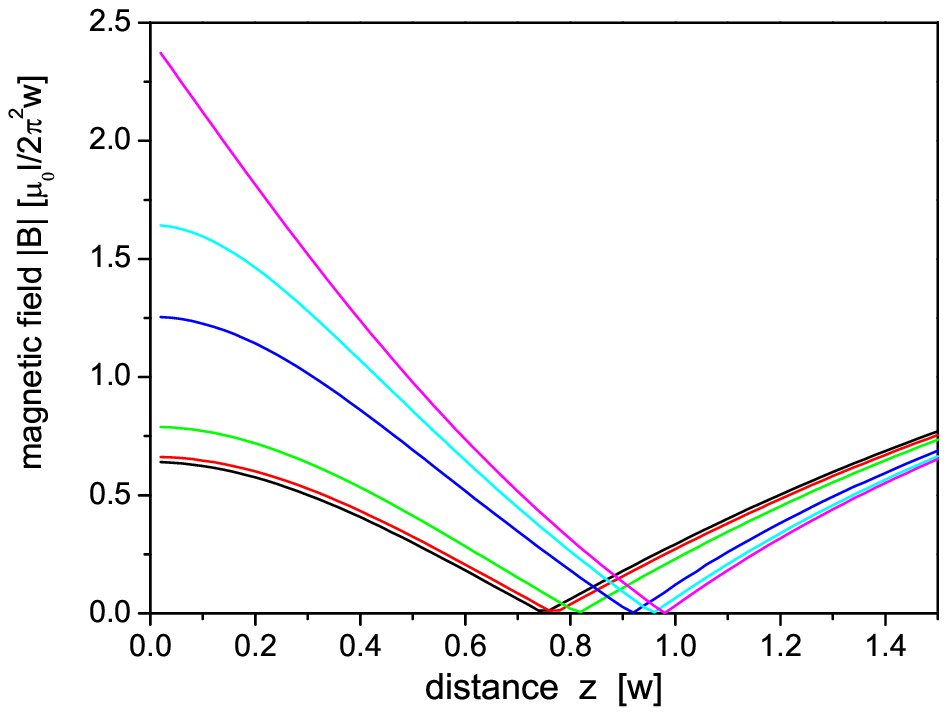}
\caption{\small{}(top left) Sheet current distribution in a thin
strip of type-II superconductor as a function of the transport
current $I$ which is increased from zero (virgin state). The
curves are plotted for $I/I_c = 0.2,\, 0.5,\, 0.85,$ and $0.95$
where $I_{c} = 2 d w j_{c}$ is the (total) critical current;
(top right)  Magnetic field vs.\ distance and current above the centre
of the strip, $x = 0$, with no external (bias) field.
For a critical current density of $j_{c} = 10^{11} \, {\rm A/m}^2$
the field reaches a maximum value of $\mu_{0}
I_{c}/(2\pi^{2}w)\approx127\,{\rm G} (d/1\,\mu{\rm m})$ at
$I = I_{c}$ and at the wire surface. 
(bottom left) Magnetic fields calculated as a
function of distance for different $I/I_c$ ratios, as given in the 
legend: black -- Meissner state;
red \ldots cyan -- $I / I_{c} = 0.2 \ldots 0.95$;
magneta -- normal metal. The field is normalized to
$\mu_{0} I / (2\pi^2 w)$, as in Fig.\ref{fig:b-strip}.
The calculations were made numerically
using the current distributions given 
by~(\ref{eq:Bean-Brandt-current-density}). 
(bottom right) The ``mixed-state" magnetic
potentials as a function of distance $z$ at $x=0$ are presented for 
the same
$I/I_c$ ratios. The horizontal bias field equals $2.5\mu_{0}
I/(2\pi^{2}w)$.} \label{Fig5}
\end{figure}

Let us consider a side guide trap realized by a type II
superconducting wire in the form of a strip
(Fig.\ref{fig1:sketch-strip}), carrying a transport current $I$ in
zero external field. To calculate the current distribution in the
strip we use the Bean critical state model \cite{Bean}: the
current (area) density can only take three different values: $\pm
j_{c}$ or $0$. Following Brandt \cite{Brandt}, the sheet current
(integrated over the thickness of the strip) is determined as
$J(x) = (d_{+}-d_{-}) j_{c}$, where $d_{+}$ and $d_{-}$ are the
history-dependent thicknesses of the regions carrying $+ j_{c}$ or
$- j_{c}$, respectively ($d_{+}+d_{-} \leq d$). The sheet current
cannot be higher than its critical value $J_{c} = d j_{c}$. This
value is achieved in regions near the strip edges. In the central
part of the strip, a field-free region exists that is shielded by
the current-carrying domains from the magnetic field, similarly to
the Meissner state. In this model for a type II superconducting
strip, the distribution of the sheet current is given by the
expression \cite{Brandt}:
\begin{equation}
   J(x,J_{c},I) =
   \left\{
   \begin{array}{ll}
       (2J_{c} / \pi) \arctan\sqrt{\frac{w^{2}-b^{2}}{b^{2}-x^{2}}},
       & \left|x\right|<b
       \\
       J_{c},
       & b<\left|x\right|<w
   \end{array}
   \right.
   \label{eq:Bean-Brandt-current-density}
\end{equation}
where $b=w\left(1-I^{2}/I_{c}^{2}\right)^{1/2}$, $I_{c} = 2wJ_{c}$ is
the maximal value for a superconducting current in the strip (critical
current); it is achieved at full magnetic field penetration, $b = 0$.  The
distribution of the sheet current is presented in Fig.\ref{Fig5} (top left).
Eq.(\ref{eq:Bean-Brandt-current-density}) applies when the
transport current $I$ has been increased from zero (virgin state) for
the first time (cyclic current change is discussed below).

Due to symmetry, the side guide is located above the strip centre
when the bias field is parallel to the $x$-axis. Fig.\ref{Fig5}
(bottom left) presents the magnetic field produced by a current
$I$ on the axis $x=0$ of an infinitely thin strip of width $2w$ in
the mixed state. The dependence of the magnetic field
on the distance $z$ was calculated
numerically for four ratios of the current to the critical
current $I/I_c$ (the same as in Fig.\ref{Fig5} (top left)).
The data should be compared to Fig.\ref{fig:trap-height} where
a superconductor in the Meissner state and a normal conductor
are considered. The magnetic trapping potential that is
formed in combination with a horizontal bias field
(of magnitude $2.5\mu_{0} I/(2\pi^{2}w)$), are shown in Fig.\ref{Fig5} 
(bottom right). These plots demonstrate the tendency: the closer the
current to the critical current, the closer are the trap
parameters to the values of a normal metal trap.

It should be noted that a superconductor in the mixed state is
essentially a nonlinear system. As can be seen from the 3D-plot
presented in Fig.\ref{Fig5} (top right) the magnetic field is a
nonlinear function of the current, in contrast to both a normal
conducting wire and a superconducting one in the Meissner state.
This happens because the shape of the current distribution depends
on the ratio $I / I_{c}$ [see Fig.\ref{Fig5} (top left)].
The calculations show that for low currents ($I < 0.2 \,I_{c}$),
the nonlinearity is negligible and the field distribution around
the type~II superconducting strip may be described by the
expressions obtained for a strip in the Meissner state in Section
\ref{s2:Meissner-state}. In the opposite case, $I\cong I_{c}$, the
current density equals the critical value over almost the whole
strip width. The magnetic field around the strip can then be
calculated as for a normal metal, with a spatially uniform current
density.

\begin{figure}[tbh]
\centerline{
\includegraphics[width=90mm]{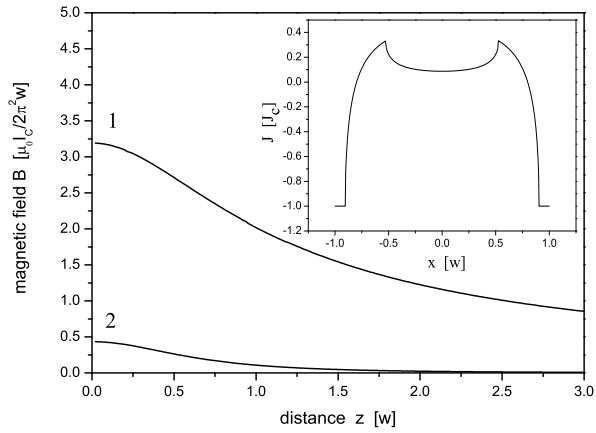}
} \caption{\small{} Illustration of trapped magnetic flux after
ramping up and down a transport current in a type II
superconducting wire in the mixed state. We plot the magnetic
field produced by a transport current $I$ = 0.85 $I_{c}$ above the
centre of a strip wire (curve~1), and the magnetic field caused by
the flux trapped in the strip, after decreasing the current again
to zero (curve~2).  The magnetic field is given in units of
$\mu_{0} I_{c}/(2\pi^{2}w)$.
The inset shows the remnant sheet current
$J( x )$ in the strip after the current cycle, normalised to
the critical sheet current $J_{c}$. The calculation follows
Ref.\cite{Brandt}.  }
\label{Fig7}
\end{figure}

\subsubsection{Trapped magnetic flux and related noise}

The capability of type II superconductors to freeze a magnetic
flux results in the irreversible behavior of the current profile
in the mixed state strip under cyclic changes of the external
fields or transport currents. Fig.\ref{Fig7} presents the magnetic
field above the superconducting strip after ramping up the
transport current to $I = 0.85 \,I_{c}$ (curve 1) and decreasing it
again to zero (curve 2).  The inset shows the inhomogeneous current
distribution that
is left in the strip as a result of the freezing of the magnetic
flux in this current cycle: in the centre, the current keeps
flowing in the direction of the peak transport current, while near
the edges it flows backwards. At short distance ($z < 0.2\, w$)
the frozen magnetic field is about 15\% of the field produced by
the maximum current. The ratio of these fields decreases
approximately like $1/z^{2}$ and is about 1\% at $z = 3\,w$. These
calculations were made for zero external (bias) field in the
framework of the Brandt model \cite{Brandt}.

The effect of the frozen magnetic field has to be taken into
account for the proper loading and control of the magnetic trap.
The frozen flux can also be the main source of magnetic noise generated
by the guiding wire because of the motion of the trapped vortices.
Other effects which can influence the atom
cloud above mixed state wires are the variations of the critical
current density due to local changes in the magnetic field or due
to structure inhomogeneities.  In addition to the vortex lattice,
these effects could corrugate the trapping potential at low trap
heights, $z_{t}\leq w$. Changes on slow time scales may occur due
to the re-distribution of the frozen flux~\cite{Bobyl,Yeshurun}.

The magnetic noise (flux noise) due to the motion of the trapped vortices 
is much higher than the noise in the Meissner state. The
flux noise was investigated in many experimental
\cite{Clem,Ferrari,Placais} and theoretical
\cite{Clem,Placais,Shapiro95} works. It was shown that it
is closely related to the mechanism of vortex pinning which
in turn depends on material properties and preparation technology
of the sample. The flux noise intensity is determined by external
conditions as well. When the temperature, current and magnetic
field are far from the critical values $T_c$, $J_c$, and $B_{c2}$,
the noise arises from thermally activated, mutually incoherent
hoppings among pinning centres.  For this regime of ``flux creep'',
SQUID-measurements near high-$T_{c}$ films demonstrate
that the noise is strongly dependent on the film quality
\cite{Ferrari}. 
In addition, the flux noise picture is more complicated:
the noise level depends on the magnetic pre-history, it follows
not only the number of trapped vortices, but also the specific
spectrum of metastable trapped states. In particular, a long-lived 
noisy state may occur in a superconducting wire after a pulsed
transport current if the current is not small compared to
the critical value \cite{Ferrari,Kuriki}.

The influence of the magnetic pre-history is obviously harmful for
controlling cold atom traps near a superconducting surface.
Further to the additional flux noise, the vortices trapped in the
superconductor produce magnetic disorder because of the arbitrary
locations of the pinning centres. 
Note that cooling of the wire through the superconducting transition
in an external magnetic field may also lead to the freezing of vortices.
SQUID studies of the remnant noise due to vortices trapped in a weak
field show that the flux noise spectral density at low frequencies
decreases linearly, as the magnetic field in which the superconducting
transition occurred, is lowered \cite{Ferrari,Shaw}.
In fact,
the remnant noise is proportional to the number of vortices
trapped in the sample in agreement with the Dutta, Dimon
and Horn model \cite{DDH}. This number is also determined by the
shape of the superconductor: the narrower the strip, the less
vortices are frozen. As was shown in Ref.\cite{Stan}, a niobium
strip of width $2w =10\,\mu{\rm m}$ placed in a vertical magnetic
field smaller than $\approx 0.5$ G does not freeze vortices while
being cooled through $T_c$. At the same time, for a 
width of 100\,$\mu{\rm m,}$ the critical field for vortex freezing
is less than 4\,mG \cite{Stan}. These observations indicate the
advantageous nature of narrow superconducting wires for atom chip
experiments which require particularly long lifetimes and
coherence times.

To summarise the possibility of using mixed state wires for
atom guiding and trapping, the advantage over the Meissner state
is the higher current and thus the tighter confinement,
while the disadvantage is the expected
static potential corrugation and higher noise. 
It may even turn out that their performances are 
comparable to other, normally conducting materials that have 
been suggested recently~\cite{Dikovsky,Tal}, but more detailed
investigations are needed here.
On the other hand, it may be possible to trap cold atoms close
enough and long enough so that they may be considered as a probe
of the mixed state of type II superconductors. 
The low-frequency flux
noise in the mixed state could be detected by the spin
dephasing rate of cold atoms, as was noted in
\cite{Scheel1}. 
The high
sensitivity of cold atoms to a disturbance of the magnetic
potential could be used for visualization of the static disorder
produced by frozen vortices, analogously to the current static
scattering in normal atom chip wires
\cite{Wildermuth05,FolmanScience,Yoni}. 
Atom clouds having high spatial resolution
(3$\,\mu$m) combined with excellent sensitivity to magnetic field
(4 nT) \cite{Wildermuth05} (or even better than that
\cite{Stamper-Kurn}) could provide complementary information about
the distribution and dynamics of vortices. According to the review
in Ref.\cite{Oshima}, such a combination of resolution and
sensitivity would be one of the best among various vortex
observation methods.

\section{Comparison between normal and superconducting magnetic traps}
\label{s4:discussion}

Our calculations show that magnetic traps can be created in atom
chips with superconducting wires. The main differences to normal
wires are the inhomogeneity of the current distribution and
nonzero screening currents induced by bias fields. The current
density in a superconducting wire is smaller in its centre
compared to a normal conductor (at the same total current). For
this reason, the magnetic field near the wire surface is weaker,
and a side guide trap (in a given parallel bias field) is closer
to the surface (see Fig.\ref{fig:trap-height}).

Let us now analyse in more detail the difference in trap
parameters between side-guide traps created by the normal and
superconducting guiding wires.  We also compare 
the Meissner and mixed states with respect to their trapping
``capabilities''. Note that we
do not take into account here the bending of the wires into ``U"-
and ``Z"-shapes for 3D traps (see, for example, 
\cite{Folman,Reichel02,Fortagh07}).
Calculation of the current density in bent superconducting wires
is complicated because it requires to solve a three-dimensional
problem even in a planar configuration~\cite{Cano08}.  
Our results can be applied
to the central part of the wires, sufficiently far away from the
bends.

Two figures of merit describing the confinement of cold atoms in
the magnetic trap are used for this comparison: the magnetic
field gradient at the trap centre and the depth of the trapping
potential. The trap depth is determined as the minimal height of
the (total) potential barrier, from the trap centre to either the
surface or away from it. Here, the gravitational potential is
taken into account. We adjust the trap height by setting the bias
field to the required value and take the same geometry and total
current for a fair comparison between superconducting and metallic
wires.

\begin{figure}[tbp]
\includegraphics [width=0.8\textwidth] {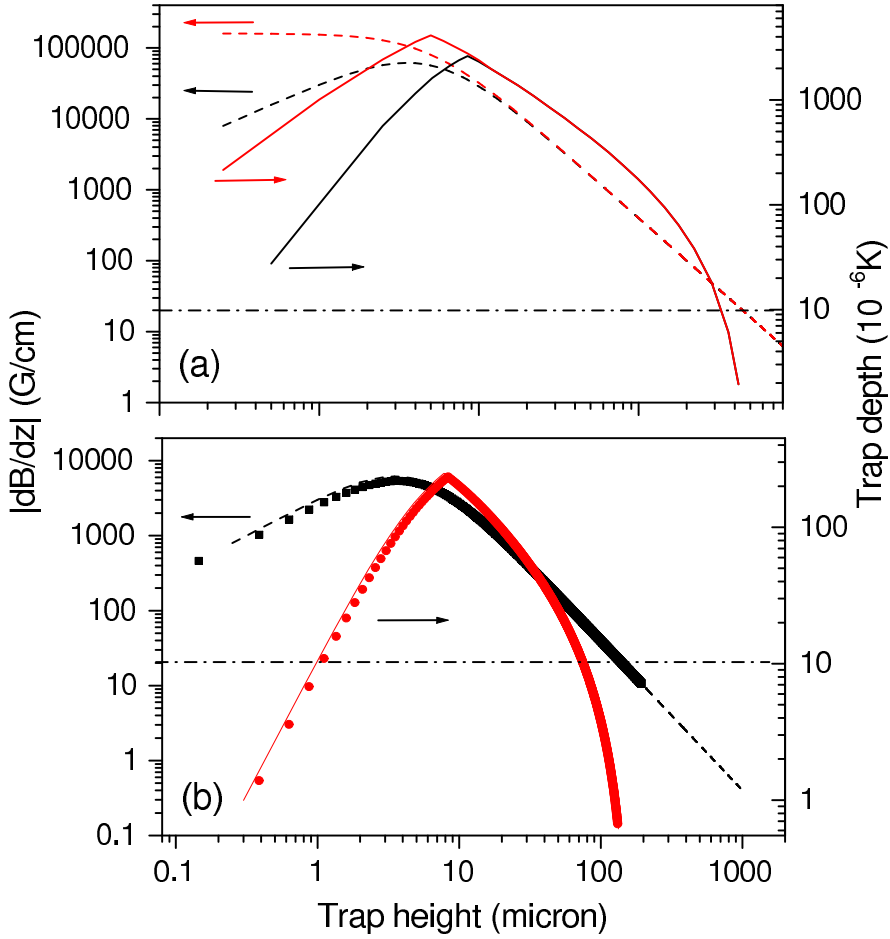}
\caption {\small{Comparison of trap parameters (field gradient and
trap depth) calculated for side-guide traps with superconducting
and normal metal wires. (a) Superconductor in the mixed state
(YBa$_{2}$Cu$_{3}$O$_{7-\delta}$ at 4.2 K) and normal metal (black
and red lines, respectively). Calculations were performed
analytically for infinitely thin guiding wires of width $2w = 10\,
\mu{\rm m}$, and current $I = 200 \,{\rm mA}Ê\approx 0.1\, I_{c}$
(assuming a thickness $d = 300\,{\rm nm}$). The trap height is
varied by adjusting the bias field. Dashed lines: absolute value
of magnetic field gradient $\left| dB / dz \right|$; solid lines:
trap depth. (b) Superconductor in the Meissner state (niobium at
4.2 K). Lines: analytical calculation for an infinitely thin wire
of width $2 w = 10 \,\mu{\rm m}$, carrying 20 mA of current. Black
and red squares: numerically determined trap parameters for a
strip with thickness $d = 2\,\mu{\rm m}$. The numerical
calculation takes into account the additional screening currents
created by the bias field (negligible for an infinitely thin
wire). Dash-dotted (horizontal) lines: minimum field gradient
required to stabilize the trap against gravity and minimum trap
depth to confine a cold atomic sample at $1\,\mu{\rm K}$. The trap depth is
calculated from the height of the potential barriers towards or
away from the chip, including gravity (``upside down'' setup with
the trap below the chip). }} \label{Fig8}
\end{figure}

Results for typical flat wires are shown in Fig.\ref{Fig8}.
Comparing superconducting and normal guiding wires, we see that
the field gradients (left scale) and the trap depths are
essentially the same for trap distances $z_{t} \ge 2\, w$, similar
to what is seen in Fig.\ref{fig:trap-height}. The difference is
maximal in the range of low trap heights, where one becomes
sensitive to the weaker magnetic field and current density in the
superconductor. This result is in qualitative agreement with the
numerical simulations of confinement parameters of specific
superconducting traps performed in \cite{Cano08}. The finite
thickness of the wire does not significantly change the picture,
as can be seen from the numerical data plotted in
Fig.\ref{Fig8}(b) (symbols). For example, an infinitely thin
superconducting strip creates a magnetic field gradient along the
$z$-direction at the trap centre (height $z_{t} = 0.75\,w$) that
is about 1.8 times less than in a normal metal trap. Along the
$x$-axis the ratio of the gradients (normal metal to
superconductor) is about 1.3.

The data presented in Fig.\ref{Fig8} are calculated for the
high-$T_{c}$ superconductor YBa$_{2}$Cu$_{3}$O$_{7-\delta}$ (YBCO)
and for niobium (Nb) in the Meissner
state. If the YBCO film is used at 4.2 K, the assumed current of
200 mA is about 10\% of the critical value (critical current
density $j_{c} \approx 7 \cdot 10^{11}$ A/m$^{2}$). Nevertheless,
we expect the wire to be in the mixed state, at least at its edges
where the current density exceeds the values permitting the
Meissner state. For niobium, the critical current density is
lower, and this is why we reduce the transport current to 20 mA.
The horizontal lines in Fig.\ref{Fig8} mark typical criteria for
reliable trapping of atoms at a temperature of $1\,\mu{\rm K}$:
the trap depth should exceed 10 $\mu{\rm K}$ and the gradient
should be high enough to protect the atoms from gravity's pull
(corresponding to $15.3\,{\rm G/cm}$ for $^{87}$Rb atoms in the
$|F = 2, m_{F} = 2\rangle$ state, where $F$ is the total spin and
$m_{F}$ its projection on the local magnetic field).

According to Fig.\ref{Fig8}(a), we predict that an atom chip based on
a YBCO superconducting strip can trap cold atoms in a wide
range of trap heights $0.2 \div 300 \,\mu{\rm m}$.
For niobium, the range is smaller ($1\div 75 \,\mu{\rm m}$) due to the
lower guiding current (Fig.\ref{Fig8}(b)).
The trap parameters are still high enough, however, to successfully
trap cold atoms, both in the Bose-Einstein condensed phase and above.
It is also seen in Fig.\ref{Fig8}
that the trap parameters in the closest vicinity of the
surface ($z \leq 5 \,\mu{\rm m}$) are worse than for a normal strip.
However, a significant gain in the lifetime due
to the reduction of magnetic noise near the superconductor
makes this trap design more attractive.

\section{Conclusion}

We have presented a theoretical analysis of superconducting atom
chips. Our methods have been both analytical and numerical; they
complement the results recently reported in Ref.\cite{Cano08}. In 
particular, the analytical expressions given here can be used for 
semi-quantitative estimates and to identify scaling laws for chip 
design.
We confirm the possibility of trapping cold atoms in a wide
range of distances ($0.2 \div 300 \,\mu{\rm m}$) with the same wire.
This analysis takes into account the specific behavior of
superconductors that carry a transport current in a magnetic
field. These peculiarities enforce modifications in the loading
procedure and the control of the atom cloud. The application of
superconductors to atom chips may enable improved atom optics with
suppressed effects of noise, as well as novel insight regarding
the noise and current distribution in superconductors.

\subsubsection*{Acknowledgments}

We acknowledge the support of the Marie-Curie programme of the
European Union, from the \emph{Bundesministerium f\"{u}r Bildung und
Forschung} (Germany, DIP project), the German-Israeli Foundation
for Scientific Research (GIF), the American-Israeli Foundation (BSF),
and the Israeli Science Foundation. C.H.\ thanks the \emph{Deutsche
Forschungsgemeinschaft} (DFG) for support (He 2849/3).

\appendix

\section  {Side guide magnetic trap based on cylindrical
superconducting
wire}
\label{a:cylinder}

\begin{figure}[tbp]
\centering
\includegraphics [width=0.50\textwidth] {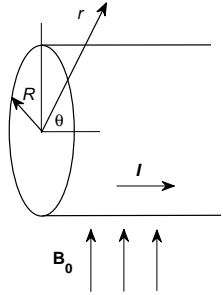}
\caption{\small{}Cylindrical wire and coordinate system.}\label{FigA1}
\end{figure}

Let us consider a conductor in the form of a cylinder (radius
$R$) with a DC transport current $I$
in an external magnetic field ${\bf B}_{0}$ (Fig.\ref{FigA1}).
A normal metal wire does not influence the external DC
field so that the total field is the superposition of the homogeneous bias
field and the field produced by the current.  The total
field is zero at $\theta = 0$
and at the trap height
\begin{equation}
     z_{t} = r_{t} - R = \frac{ \mu_{0} I }{ 2\pi B_{0} } - R
    \label{eq:cylinder-trap-height}
\end{equation}

A magnetic field cannot penetrate a superconductor in the Meissner
state. If the radius is much larger than the London penetration depth,
the magnetic field is zero inside the cylinder. The magnetic field around
the superconductor without transport current can be described by
the Laplace equation for the scalar magnetic potential $\psi$. The boundary
conditions are: the radial component of the magnetic field equals zero
at $r = R$; at $r \to \infty$, the magnetic field coincides
with the external one. The solution is
\begin{eqnarray}
    B_{r} & = & - \partial_{r} \psi =
    B_{0} \left( 1 - \frac{ R^2 }{ r^2 } \right) \sin
    \theta
    \label{eq:cylinder-radial-field}
    \\
    B_{\theta} & = & - \frac{ 1 }{ r } \partial_{\theta} \psi =
    B_{0} \left( 1 + \frac{ R^2 }{ r^2 } \right) \cos \theta
    \label{eq:cylinder-azimuthal-field}
\end{eqnarray}
where $B_{r}$ and $B_{\theta}$ are the radial and azimuthal components
of the magnetic field.  The magnetic field configuration around the
superconductor thus differs from a normal metal
even if a transport current is absent.  Due to the Meissner effect, the
magnetic field near the superconductor at $\theta = 0$
is increased.

\begin{figure}[tbp]
\includegraphics[height=55mm]{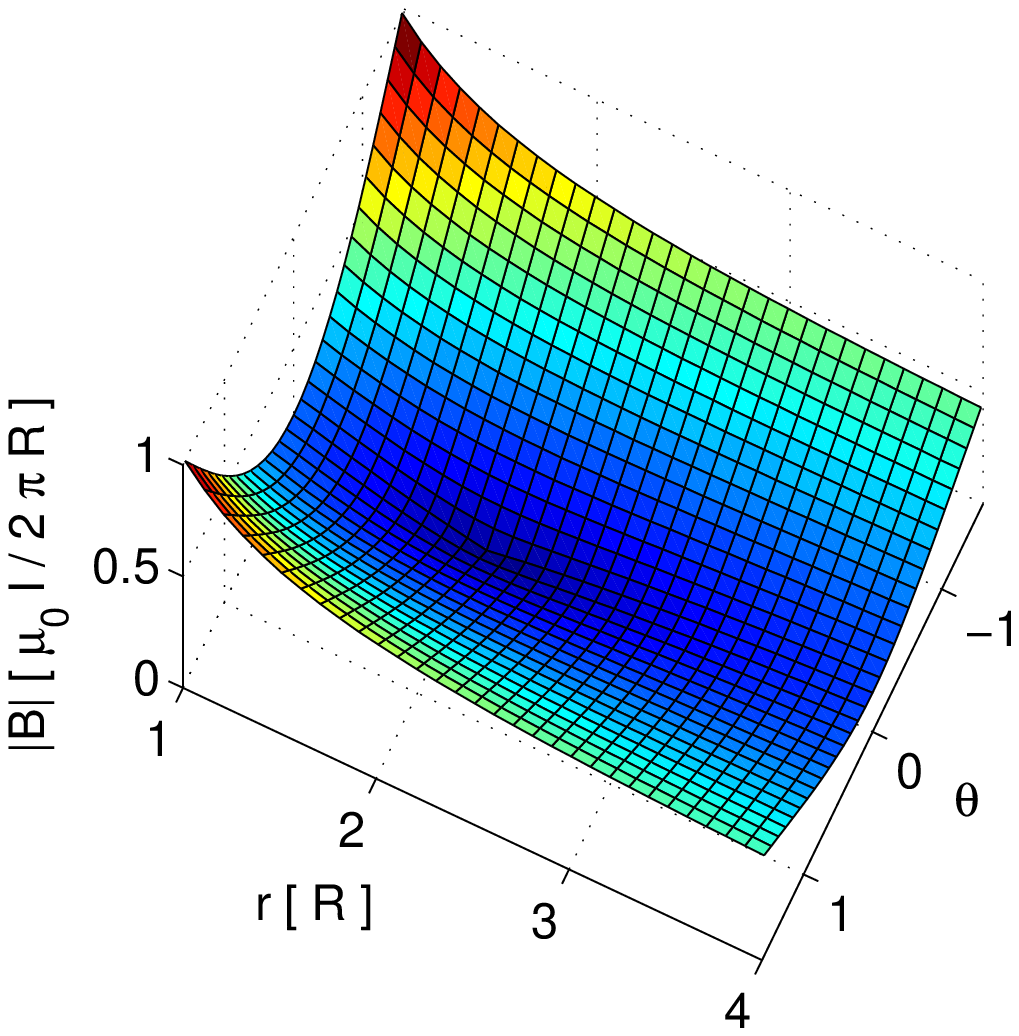}
\includegraphics[height=55mm]{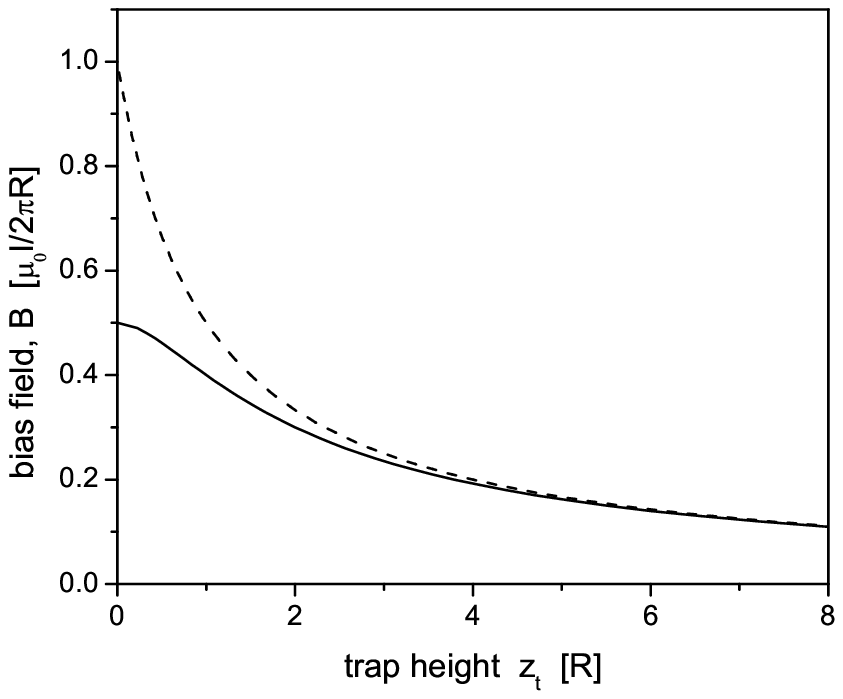}
\caption{\small{}(left) Modulus of the total magnetic field around
a superconducting cylinder with a transport current $I$ in a bias
magnetic field $B_{0} = 0.4 \times \mu_{0} I / 2\pi R$. The field
minimum is located at the point $\theta=0$,  $z_{t} = 0.8\, R$ ($r
= 1.8\,R$). (right) Required bias field as a function of the trap
height for a cylindrical wire, based on
Eqs.(\ref{eq:cylinder-trap-height})
and~(\ref{eq:cylinder-with-current}): solid line --
superconductor; dashed line -- normal metal.  In both graphs, the
magnetic field
is normalised
to $\mu_{0} I / 2\pi R = 2\, {\rm G} (I / 1\,{\rm A} ) / (R / 1\,{\rm
mm})$.}
\label{FigA2}
\end{figure}

If the cylinder carries a transport current, the outer field is the 
same as for a normal conductor, due to the cylindrical symmetry. 
We find that 
the minimum of the total field
(Fig.\ref{FigA2}(left)) is located at $\theta = 0$
and distance
\begin{equation}
     z_{t} = r_{t} - R = R \frac{ 1 + \sqrt{ 1 - 4 h^2 } }{ 2 h } - R,
     \qquad
     h = \frac{ 2 \pi R B_{0} }{ \mu_{0} I }
        \label{eq:cylinder-with-current}
\end{equation}
from the wire surface.
Because of the screening of the bias field by a superconductor,
the same trap height in
a superconducting chip is achieved with a lower bias field
than in a normal conducting chip
(see Fig.\ref{FigA2}(right) and
also Fig.\ref{fig:trap-height}(right) for a rectangular wire).
This reduction reaches 50\% at small trap heights.
The difference decreases with an increase of
this height and practically vanishes at $z_{t}>3R$.

\section {Numerical calculation of magnetic field around a finite
thickness superconductor}

We consider a wire that is infinitely long in the
$y$-direction
and ignore here boundary effects (these become relevant for ``U''- and
``Z''-shaped wires, of course).  By symmetry, the magnetic field is
independent of $y$ and lies in the $xz$-plane (see
Fig.\ref{fig1:sketch-strip}).  Inside the superconductor, the field is
zero.  Outside, we describe the field in terms of a scalar potential
$\psi$ and a vector potential ${\bf A} = {\bf e}_y A$
\begin{equation}
{\bf B}({\bf x}) = - \nabla \psi ({\bf x}) +
\nabla \times {\bf A}({\bf x}) \; .
\label{eq:B2D}
\end{equation}
We shall see below that the scalar potential describes the field
caused by an external magnetic field (as in Appendix A), while the
vector potential gives the field caused by a transport current in the
superconducting wire.  The introduction of two potentials may seem
superfluous, because one could work, outside the wire, with the scalar
potential only.  But $\psi$ would then become a multivalued function
for a nonzero transport current.

The magnetic field normal to the superconductor surface is
continuous and must therefore vanish. (This corresponds to the radial
derivative of $\psi$ in Appendix A.)
Writing ${\bf n}( {\bf x}_{s} )$
for the normal at surface point ${\bf x}_{s}$, we have
\begin{eqnarray}
{\bf n}( {\bf x}_{s} ) \cdot {\bf B}( {\bf x}_{s} ) & = &
-{\bf n}( {\bf x}_{s} ) \cdot
\nabla \psi({\bf x}_{s})
- ({\bf n}( {\bf x}_{s} ) \times {\bf e}_{y} )
   \cdot \nabla  A({\bf x}_{s})
   \nonumber\\
& =& -\frac {\partial \psi({\bf x}_{s})} {\partial n({\bf x}_{s}) } -
\frac{\partial  A({\bf x}_{s})}{\partial t ({\bf x}_{s}) }
= 0  \;
 \label{eq:surf}
\end{eqnarray}
where ${\partial} / {\partial n}( {\bf x}_{s} )$
and ${\partial} / {\partial t}( {\bf x}_{s} )$
are the normal and tangential derivatives at ${\bf x}_{s}$.
(The local tangent vector is ${\bf t} = {\bf n} \times {\bf e}_{y}$.).
We construct the potentials $\psi$ and $A$ such that both terms
${\partial \psi _{s}} / {\partial n}$ and ${\partial  A_{s}} /
{\partial t}$ in Eq.(\ref{eq:surf}) are zero. Note that
${\partial  A_{s}} / {\partial t} = 0 $ indicates that $A( {\bf
x}_{s} )$ is
a constant on the surface which we denote by $A_{0}$.  The tangential
magnetic field is nonzero at the (outer)
surface of the superconductor; the interior of the wire is screened
from this field by a surface current
(along the $y$-axis) of magnitude ${\bf t} \cdot {\bf B} / \mu_{0}$.

Considering that outside the wire, both divergence and curl of the
magnetic field are zero, we get
\begin{equation}
\nabla ^2 \psi ({\bf r}) = 0
, \qquad
\nabla ^2 A ({\bf r}) = 0
\label{eq:divergence}
\label{eq:curl}
\end{equation}
where  $\nabla ^2$ is the two-dimensional Laplace operator. These
Laplace
equations can be solved with the help of the Green function
\begin{equation}
G({\bf x}, {\bf r}) = -\frac {1}{2\pi }\log r \; ,
\label{eq:shortgreen}
\end{equation}
where $r=|{\bf r} - {\bf x}|$, given the values of the potentials and
their derivatives on the wire surface, and an asymptotic condition at
large distance.

The two potentials behave differently with respect to
the current $I$ transported by the superconductor. It can be seen by
recalling the Amp{\` e}re-Maxwell law (in the static limit)
 \begin{equation}
\mu_{0} I = \oint_{S} {\bf B}({\bf x}) \cdot {\rm d}{\bf a}
 = - \oint_{S} \frac{ \partial A }{ \partial n }  {\rm d}{a} \; ,
 \label{eq:A}
 \end{equation}
where $I$ is the total  current flowing through the superconductor,
$S$ a closed curve including the cross-section of the superconductor
(with oriented line element ${\rm d}{\bf a}$), and $\partial /
\partial n$
is the derivative normal to  the curve $S$ and pointing `outside'.
Hence, the vector potential $A$ is proportional to the transport
current.
Conversely, the scalar potential asymptotically goes over into
$\psi \to \psi_{\rm ext}( {\bf r} ) = - {\bf r} \cdot {\bf B}_0$ where
${\bf B}_0$ is the homogeneous bias field. This asymptotic condition
forces
$\psi$ to be proportional to ${\bf B}_0$.

We solve the Laplace equations~(\ref{eq:curl}) for the potentials
$M = \psi, A$ in terms of a surface integral equation (${\bf r}$
outside the wire)~\cite{Jackson,Vesperinas}
\begin{equation}
M({\bf r})  =  M_{\rm ext}({\bf r}) - \oint _{S}{\rm d}a({\bf x}) \left(
G({\bf x},{\bf r}) \frac{\partial M({\bf x})}{\partial n({\bf x})} -
\frac{\partial G({\bf x},{\bf r})}{\partial n({\bf x})} M ({\bf
x})\right)
      \; ,
\label{eq:M}
\end{equation}
where
$M_{\rm ext}$ is the external
potential (nonzero only for $\psi$),
$S$ is now the circumference of the wire (with scalar line element
${\rm d}a({\bf x})$), and
the normal derivative
${\partial} / {\partial n({\bf x})}$ points `outside' the wire.
For $M=\psi$, we have ${\partial \psi({\bf x})} / {\partial n({\bf
x})}
= 0$ on the surface according to Eq.(\ref{eq:surf}); then there is
only
one unknown, $\psi( {\bf x} )$, in Eq.(\ref{eq:M}). When $M=A$,
we can set under the integral $A( {\bf x} ) = A_{0}$, constant on
the surface, also according to Eq.(\ref{eq:surf}). Then the integral
of the second term in the bracket can be shown to vanish, and
there is only
one unknown left, ${\partial A({\bf x})} / {\partial n({\bf x})}$, which is
actually the surface current density, see Eq.(\ref{eq:A}).
    We always choose $A_0 = 1$ in the beginning, compute the total
current
    by Eq.(\ref{eq:A}) (giving a coefficient like the wire inductance)
    and re-scale $A( {\bf x} )$ and $\partial A/ \partial n( {\bf x} )$ 
    to get the desired current.

The potential $M$ and its derivative  on the
surface are obtained by letting ${\bf r} \to {\bf x}_{s}$
in Eq.(\ref{eq:M}) approach the surface. Note
that we touch the singularity of the Green function
$G({\bf x},{\bf x}_{s})$ and
its normal derivative
${\partial G({\bf x},{\bf x}_{s})} / {\partial n({\bf x})}$
under the integral when ${\bf x}={\bf x}_{s}$. To handle this, we
expand the integrand within a small neighborhood $s$ of length ${\rm
d}a$
around ${\bf x}_{s}$ and perform the integration,
giving
\begin{equation}
    \int_{s}\!{\rm d}a( {\bf x} )
    G( {\bf x}, {\bf x}_{s} )
    \frac {\partial M({\bf x})}{\partial n( {\bf x} ) }
    \approx \frac{1}{2\pi}
    \frac{\partial M({\bf x}_{s})}{\partial n( {\bf x}_{s} ) }
    \left ( 1 - \log \frac{{\rm d}a}{2} \right)  {\rm d}a \, .
\label{eq:log-singularity}
\end{equation}
\begin{equation}
\int_{s}\!{{\rm d}a}( {\bf x} )
\frac{\partial G({\bf x}_{s},{\bf x})}{\partial n({\bf x})}
M ({\bf x}) \approx
\left( \frac{1}{2}+\frac{{\rm d} \phi( {\bf x}_{s} ) }{4\pi}
\right)
M ({\bf x}_{s}) \; .
\label{eq:tsing}
\end{equation}
Here the angle ${\rm d} \phi( {\bf x}_{s} )$ involves the radius of
curvature $R({\bf x}_{s})$ of the surface:
${\rm d} \phi( {\bf x}_{s} ) = {\rm d}a / R({\bf
x}_{s})$, it describes the angle subtended by the surface element
as seen from the centre of curvature. The logarithmic correction of
Eq.(\ref{eq:log-singularity}) and the curvature
correction~(\ref{eq:tsing})
greatly improve the convergence of the numerical calculations.

The other parts of the surface integral, excluding the point ${\bf
x}_{s}$, are discretized in the usual way, mapping the integral
equation into a linear system~\cite{Rogobete04a,Bancroft}. For
example, a rectangle 20.64 $\mu {\rm m} \times$ 0.84 $\mu {\rm m}$
with rounded corners ($R = 0.32 \mu {\rm m}$) is typically discretized
into $420$ surface elements along its circumference.  This leads to a
linear system with $420^2$ matrix elements.  Once we have the
potential or its derivative on the wire surface,
the field outside the wire is found from Eq.(\ref{eq:M}). We have
checked the convergence of the numerics, for different
discretizations of the wire surface, against the exact solution of
Appendix A for cylindrical wires and against the solution of
Ref.\cite{Brandt00} for rectangular wires with sharp corners.

\end{document}